%% file: main.tex
\documentclass[aps, prd, twocolumn, groupedaddress, nofootinbib]{revtex4-2}

\usepackage{amsmath}
\usepackage{amssymb}
\usepackage{bbold}
\usepackage{graphicx}
\usepackage{mathtools} 
\usepackage{lipsum}
\usepackage{xcolor}
\usepackage[hidelinks, colorlinks=true, citecolor=blue, linkcolor=black, urlcolor=blue]{hyperref}
\usepackage{cleveref}
\usepackage{booktabs}               % Tabellen
\usepackage{subcaption}

\newcommand{\iu}{\mathrm{i}}
\newcommand{\eu}{\mathrm{e}}
\newcommand{\dd}{\mathrm{d}}

\renewcommand{\vec}[1]{\mathbf{#1}} % exchange vec with bold-math letter
\begin{document}

% \preprint{APS/123-QED}

\title{Energy-momentum response to metric perturbations in the fluid dynamic regime}

\author{Tim Stoetzel}
\email[]{tim.stoetzel@uni-jena.de}
\affiliation{Theoretisch Physikalisches Institut, Friedrich-Schiller-Universität Jena, Max-Wien-Platz 1, 07743 Jena, Germany}

\author{Rebekka Fechtner}
\email[]{rebekka.fechtner@stud.uni-heidelberg.de}
\affiliation{Theoretisch Physikalisches Institut, Friedrich-Schiller-Universität Jena, Max-Wien-Platz 1, 07743 Jena, Germany}

\author{Stefan Floerchinger}
\email[]{stefan.floerchinger@uni-jena.de}
\affiliation{Theoretisch Physikalisches Institut, Friedrich-Schiller-Universität Jena, Max-Wien-Platz 1, 07743 Jena, Germany}

\begin{abstract}
The interplay of relativistic fluid dynamics and spacetime geometry is discussed in the regime of small wave numbers and frequencies. 
A combination of gravitational Ward identities and fluid dynamic equations of motion in the  Mueller-Israel-Stewart formulation is used to explicitly determine the retarded linear response of the energy-momentum tensor to metric perturbations. 
We also discuss applications to gravitational wave production and the damping of gravitational waves in a relativistic fluid. 
\end{abstract}

\maketitle

\section{Introduction}
\input{inputs/introduction.tex}

\section{Linear response}
\label{sec:Linearresponse}
\input{inputs/linearresponse.tex}

\section{Relativistic fluid dynamics}
\label{sec:Relativisticfluiddynamics}
\input{inputs/hydro.tex}

\section{Calculation of the response function}
\label{sec:Calculationoftheresponsefunction}
\input{inputs/calculations.tex} 

\section{Production and absorption processes close to equilibrium}
\label{sec:Productionandabsorptionprocessesclosetoequilibrium}
\input{inputs/prod_and_decay.tex} 

\section{Conclusion}
\label{sec:Conclusions}
\input{inputs/conclusion.tex}
\subsection*{Acknowledgments} 
We thank Christian Schmidt for some helpful discussions on the topic of gravitational perturbations on time-dependent backgrounds and comments on the manuscript.
We would like to acknowledge the use of perceptional uniform colour maps, defined in ref~\cite{crameri_2023_8409685}.  

\appendix 
\input{inputs/appendix.tex}
\newpage
\bibliography{main.bib}

\end{document}

%% file: inputs/introduction.tex
One of the most interesting questions for non-equilibrium quantum field theory is how quantum fields interact with the spacetime metric $g_{\mu\nu}(x)$. 
This is of relevance for cosmology or astrophysical situations, but also presents an interesting conceptual question by itself. 

While quantum field theory in Minkowski space can often be restricted to excitations of a vacuum state or to global thermal equilibrium, non-equilibrium states occur generically in curved spacetime~\cite{Calzetta:2008iqa,Wald1995,Parker:2009uva}. 
In fact, for non-stationary situations, or in the absence of a global time-like Killing vector field, there are actually no global equilibrium states~\cite{Tolman:1930ona}. 
Of course, non-equilibrium situations can arise in many ways, for example from time-dependent external fields or as an initial value problem~\cite{Calzetta:2008iqa,Kamenev_2011,Berges:2004,Parker1969,mukhanov_winitzki_2007,Hawking1975}. 
However, the non-equilibrium states that are generated through a time-dependent spacetime metric are particularly interesting. 
This is related to the fact that all matter fields couple to the metric, and in a very universal way. 
For example, one could have a metric that is first stationary and start with a thermal equilibrium state, before the metric becomes time-dependent and takes the state out-of equilibrium. 
One may then ask how the quantum field theory responds, for example how field expectation values or correlation functions depend on the time-dependent metric perturbation. 
This setup encompasses many interesting problems in cosmology, but also allows determining the reaction of quantum fields to a gravitational wave passing by.

When the energy-momentum tensor is followed in terms of its expectation value and correlation functions one can also address questions about back-reaction of matter fields on the metric -- at least approximately when these perturbations are small~\cite{Calzetta:1986ey,DeWitt:1975ys,birrell_davies_1982}. 
This touches eventually fascinating questions concerning the interplay of gravitational theory and quantum field theory~\cite{mukhanov_winitzki_2007, birrell_davies_1982, Wald1995}.
For example the dissipation of gravitational radiation by matter on cosmological scales has been studied in refs~\cite{Hawking:1966qi,Flauger:2017ged,Weinberg:2008zzc}. 
Other interesting applications include the study of gravitational wave background from the early universe~\cite{Ghiglieri:2015nfa,Ghiglieri:2022rfp}, constraints on inflationary models~\cite{Abbott:1984fp,Caprini:2018mtu} or studying possible phase transitions in the first second of the universe~\cite{Mazumdar_2019,Caprini:2018mtu}. 

In the present work we aim to discuss the interplay of quantum field theory and the spacetime metric in a regime where the quantum field theory is strongly interacting so that its typical relaxation time for the approach to local thermal equilibrium is small compared to the time scales where the spacetime metric is varied. 
One would believe that quantum fields can be described as a \textit{fluid} in this regime, based on an extension of concepts from thermodynamics to a notion of local thermal equilibrium~\cite{Calzetta:2008iqa,annurev.nucl.54.070103.181236,doi:10.1142/S0217751X13400113}. 
In a situation where there are no conserved quantum numbers besides energy and momentum, relativistic fluid dynamics is essentially a theory for the energy-momentum tensor and therefore describes very naturally the interplay of quantum fields with metric perturbations.

We shall concentrate on situations where the starting point is a global thermal equilibrium state, and where the metric perturbations are weak so that the powerful formalism of linear response theory~\cite{Kubo:1957mj, Kubo:1957wcy} can be applied. 
We will assume that a relativistic fluid description is valid, and work out the different response and correlation functions for the energy-momentum tensor in this situation. 
In the linear response setup this allows to find an essentially complete characterization of the perturbation in the energy-momentum tensor to metric perturbations.

The linear response is usually relevant when determining transport coefficients, which then can be calculated via Kubo formulas~\cite{Kadanoff:1963axw,Calzetta:1999ps,Jeon:1994if}.
Derivations of these in the context of relativistic fluid dynamics have been done in refs~\cite{Moore:2010bu,Czajka:2017} in the first- and second-order formulation.
A recent work~\cite{Jeon:2025} does investigate the general analytic structure of fluid dynamical response to gravitational perturbations, also considering second- and additionally third-order fluid dynamical theories. 
In their work, the general analytic structure of these theories was investigated in four spacetime dimensions and new Kubo formulas could be found where the limits of frequency and spatial momentum can be exchanged. 
While we do also investigate the response function of second order Mueller-Israel-Stewart fluid dynamics, we explicitly calculate the response for general spacetime dimensions. 

This paper is organized as follows.
In section \ref{sec:Linearresponse} we provide a brief formulation of linear response theory for our setup, concentrating on perturbations of the energy-momentum tensor resulting from metric perturbations and derive the Ward identity in general coordinates.
Section \ref{sec:Relativisticfluiddynamics} recalls the most important elements of causal dissipative relativistic fluid dynamics in the formulation proposed by Mueller~\cite{Mueller1967} and Israel and Stewart~\cite{Israel:1979wp} and the resulting linearized equations of motion. 
This also includes a brief discussion of ideal fluid dynamics on curved spacetime and susceptibilities related to the energy-momentum tensor.
In section \ref{sec:Calculationoftheresponsefunction} we discuss our concrete calculation of the response functions in the fluid dynamic limit, with an inversion problem constituting the main technical difficulty. 
In section \ref{sec:Productionandabsorptionprocessesclosetoequilibrium} we discuss how our results allow determining the production rate for gravitational waves from a fluid in a close-to-thermal-equilibrium situation. 
Similarly, we also discuss there how the linear response properties constrain the damping of gravitational waves that propagate through a relativistic fluid.%, which is discussed in refs~\cite{Hawking:1966qi,Weinberg:2008zzc} using the 'first order' hydrodynamics formulation. 
Finally, we draw some conclusions in section \ref{sec:Conclusions}. 
Technical details are provided in the appendix. 

\textit{Conventions}. We work with a mainly plus convention for the metric signature, such that the Minkowski metric is $\eta_{\mu\nu} = \text{diag}(-1, +1, \ldots, +1)$ in $d$ spacetime dimensions. 
Furthermore, we work in units where $c=k_\mathrm{B}=\hbar=1$.

%% file: inputs/linearresponse.tex
We discuss here the general setup of response theory for the energy-momentum tensor.

\subsection{Energy-momentum response to metric perturbations}
The starting point is a definite quantum state, usually prepared in the infinite past $t\to -\infty$. 
Typically, it is assumed that this is a global thermal equilibrium state, but that assumption is not strictly necessary. 
This state evolves then in a given spacetime geometry, with fixed coordinates $x^\mu$ and spacetime metric $g_{\mu\nu}(x)$. 
We are mainly interested in the expectation value of the energy-momentum tensor $T^{\mu\nu}(x)$ and in the question of how it depends on the spacetime geometry described with the metric $g_{\mu\nu}(x)$.

More specifically, we decompose the metric linearly,
\begin{equation}
    g_{\mu\nu}(x) = \bar g_{\mu\nu}(x) + \delta g_{\mu\nu}(x),
    \label{eq:Splittinggmunu}
\end{equation}
into a background part $\bar g_{\mu\nu}(x)$ and a small perturbation $\delta g_{\mu\nu}(x)$. 
The latter induces a small perturbation in the energy-momentum tensor,
\begin{equation}
  \mathfrak{T}^{\mu\nu}(x) = \bar{\mathfrak{T}}^{\mu\nu}(x) + \delta\mathfrak{T}^{\mu\nu}(x).
  \label{eq:SplittingTFrak} 
\end{equation}
For technical reasons we have done the decomposition for the energy-momentum tensor density 
\begin{equation}
  \mathfrak{T}^{\mu\nu}(x) = \sqrt{g(x)} T^{\mu\nu}(x),
\label{eq:defFrakT}
\end{equation}
where $g(x) = - \det g_{\mu\nu}(x)$ is the metric determinant.

We define the retarded response function as the functional derivative 
\begin{equation}
    G_\mathrm{R}^{\mu\nu\rho\sigma}(x,y) = \frac{2}{\sqrt{g(x)}\sqrt{g(y)}}\frac{\delta }{\delta g_{\rho\sigma}(y)}\mathfrak{T}^{\mu\nu}(x).
    \label{eq:response_function_definition}
\end{equation} 
The definitions are such that $G_\mathrm{R}^{\mu\nu\rho\sigma}(x,y)$ transforms under general coordinate transformations as a bi-local tensor field with indices $\mu$, $\nu$ related to the coordinate $x$ and  indices $\rho$, $\sigma$ related to the coordinate $y$. 
Relativistic causality implies that $G_\mathrm{R}^{\mu\nu\rho\sigma}(x,y)$ vanishes whenever $x$ is not in the future light-cone of $y$. 
The right-hand side of \cref{eq:response_function_definition} is typically evaluated at the background configuration $g_{\mu\nu}(x) = \bar g_{\mu\nu}(x)$.
A similar analysis of one- and two-point correlators has also been done in ref.~\cite{Osborn:1999az}.

The definition \eqref{eq:response_function_definition} then implies a linear relation between energy-momentum perturbation and metric perturbation,
\begin{equation}
  \frac{2}{\sqrt{g(x)}}\delta \mathfrak{T}^{\mu\nu}(x) = \int \dd^{d}y \sqrt{g(y)} \, G_\mathrm{R}^{\mu\nu\rho\sigma}(x, y)\delta g_{\rho\sigma}(y),
    \label{eq:linear_response_definition}
\end{equation}
in a $d$-dimensional spacetime.

\subsection{Ward identities related to diffeomorphisms}
It is useful to study the implications of invariance with respect to general coordinate transformations, or diffeomorphisms, and the closely connected covariant conservation law for the energy-momentum tensor in more detail. The latter reads
\begin{equation}
\begin{split}
    \nabla_\mu T^{\mu\nu} = & \frac{1}{\sqrt{g}}\partial_\mu \left( \sqrt{g} T^{\mu\nu} \right) + \Gamma^\nu_{\mu\rho} T^{\mu\rho} = 0,
\end{split}
\label{eq:covariant_conservation_tmunu}
\end{equation}
with the Levi-Civita connection $\nabla_\mu$.
With the splitting in \cref{eq:Splittinggmunu,eq:SplittingTFrak} we obtain
\begin{equation}
  \partial_\mu \delta \mathfrak{T}^{\mu\nu}(x) + \bar\Gamma^\nu_{\mu\rho}(x) \delta \mathfrak{T}^{\mu\rho}(x) + \bar{\mathfrak{T}}^{\mu\rho}(x) \delta \Gamma^\nu_{\mu\rho}(x)  = 0.
    \label{eq:conservation_first_order_curved_spacetime}
\end{equation}
Here one can use
\begin{equation}
\begin{split}
    \delta \Gamma^\nu_{\mu\rho} =& \frac{1}{2} \bar g^{\nu\lambda} \left( \nabla_\mu \delta g_{\rho\lambda} + \nabla_\rho \delta g_{\mu\lambda} - \nabla_\lambda \delta g_{\mu\rho} \right), \\
    = &\bar g^{\nu\lambda} \left( \frac{1}{2} \partial_\mu \delta g_{\rho\lambda} + \frac{1}{2} \partial_\rho \delta g_{\mu\lambda} -\right.\\
    &\qquad\; \left.\frac{1}{2} \partial_\lambda \delta g_{\mu\rho} - \bar \Gamma^\sigma_{\mu\rho} \delta g_{\lambda\sigma}\right),
\end{split}
\label{eq:delta_christoffel}
\end{equation}
so that one obtains an identity
\begin{equation}
\begin{split}
    & \partial_\mu \delta\mathfrak{T}^{\mu\nu} + \bar\Gamma^\nu_{\mu\rho} \delta \mathfrak{T}^{\mu\rho} \\
    & +\frac{1}{2} {\Big [} \bar{\mathfrak{T}}^{\mu\rho} \bar g^{\nu\sigma} \partial_\mu + \bar{\mathfrak{T}}^{\mu\sigma} \bar g^{\nu\rho} \partial_\mu - \bar{\mathfrak{T}}^{\rho\sigma} \bar g^{\nu\lambda} \partial_\lambda \\
    & \quad\quad - \bar g^{\nu\rho} \bar{\mathfrak{T}}^{\mu\xi} \bar \Gamma^\sigma_{\mu\xi} - \bar g^{\nu\sigma} \bar{\mathfrak{T}}^{\mu\xi} \bar \Gamma^\rho_{\mu\xi} {\Big ]} \delta g_{\rho\sigma} = 0.
\end{split}
\end{equation}
Eq.~\eqref{eq:linear_response_definition} can be utilized to express also $\delta\mathfrak{T}^{\mu\nu}$ in terms of $\delta g_{\rho\sigma}$. The resulting identity is valid for arbitrary metric perturbations $\delta g_{\rho\sigma}$. 
One finds thus the \text{Ward identity} for the retarded two-point response function~\cite{Deser:1967zzf,Herzog:2009xv}
\begin{equation}
    \begin{split}
        & \frac{\partial}{\partial x^\mu} G_\mathrm{R}^{\mu\nu\rho\sigma}(x, y) +   \bar \Gamma^\mu_{\mu\lambda}(x)   G_\mathrm{R}^{\lambda\nu\rho\sigma}(x, y) \\ 
        & +\bar \Gamma^\nu_{\mu\lambda}(x) G_\mathrm{R}^{\mu\lambda\rho\sigma}(x, y) + \left[ \bar T^{\mu\rho}(x) \bar g^{\nu\sigma}(x)  \right. \\ 
        & +\bar T^{\mu\sigma}(x) \left.\bar g^{\nu\rho}(x) - \bar T^{\rho\sigma}(x) \bar g^{\nu\mu}(x) \right] \frac{\partial}{\partial x^\mu} \frac{\delta^{(d)}(x-y)}{\sqrt{\bar g(y)}} \\
        & -\bar T^{\mu\xi} \left[ \bar g^{\nu\rho}(x) \bar \Gamma^\sigma_{\mu\xi}(x) + \bar g^{\nu\sigma}(x) \bar \Gamma^\rho_{\mu\xi}(x) \right] \frac{\delta^{(d)}(x-y)}{\sqrt{\bar g(y)}} = 0.
    \end{split}
\label{eq:ward_id_general_coordinates}
\end{equation}

\subsection{Perturbations around global equilibrium states in Minkowski space}
We now choose the expansion point to be a homogeneous configuration with a fluid in global thermal equilibrium.
The background fluid is described by the energy-momentum tensor 
\begin{equation}
    \bar T^{\mu\nu} = \bar{\epsilon} \bar u^\mu  \bar u^\nu + \bar{p}\bar \Delta^{\mu\nu},
    \label{eq:ideal_fluid}
\end{equation}
with the fluid velocity $\bar u^\mu$, and the projection orthogonal to the background fluid velocity $\bar \Delta^{\mu\nu} = \bar u^\mu \bar u^\nu + \bar g^{\mu\nu}$. 
The pressure $\bar p$ and energy density $\bar \epsilon$ are related by a thermodynamic equation of state.
We will typically work in the background fluid rest frame where $\bar u^\mu$ points in time direction. 

The perturbation $\delta T^{\mu\nu}(x)$ is taken to be caused by the deviation in the space-time metric $\delta g_{\mu\nu}(x)$ from the Minkowski configuration, $\bar g_{\mu\nu}(x) = \eta_{\mu\nu}$ \footnote{We note that Einsteins equations are \textit{not} fulfilled. Spacetime curvature in the background configuration vanishes, while the energy-momentum tensor is non-zero. In other words, we neglect the gravitational coupling $G_\mathrm{N}$.}. 
In the regime where the metric perturbation is small enough one can assume a linear relation between $\delta T^{\mu\nu}(x)$ and $\delta g_{\rho\sigma}(y)$ that is formalized by linear response theory.

Having a Minkowski background metric it is beneficial to introduce a Fourier representation,
\begin{equation}
\begin{split}
    \delta T^{\mu\nu}(t, \mathbf{x}) &= \int \frac{\dd\omega \dd^{d-1} k}{(2\pi)^d} \eu^{-\iu\omega t + \iu \mathbf{k} \mathbf{x} } \; \delta T^{\mu\nu}(\omega,\vec{k}), \\
    \delta g_{\mu\nu}(t, \mathbf{x}) &= \int \frac{\dd\omega \dd^{d-1} k}{(2\pi)^d} \eu^{-\iu\omega t + \iu \mathbf{k} \mathbf{x} } \; \delta g^{\mu\nu}(\omega,\vec{k}).
\end{split}
\end{equation}

Because of translational invariance of the background configuration, the response function defined by \cref{eq:response_function_definition} depends only on the coordinate difference, which gives rise to the Fourier representation
\begin{equation}
    \begin{split}
        &G_\mathrm{R}^{\mu\nu\rho\sigma}(x-y) = \\
        &\qquad\;\int \frac{\dd\omega \dd^{d-1} k}{(2\pi)^d} \eu^{-\iu\omega (x^0-y^0) + \iu \mathbf{k} (\vec{x}-\vec{y}) } \; G_\mathrm{R}^{\mu\nu\rho\sigma}(\omega,\vec{k}).
    \end{split}
\end{equation}
Therefore, the linear relation \eqref{eq:linear_response_definition} becomes in momentum space 
\begin{equation}
 \delta \mathfrak{T}^{\mu\nu}(\omega,\vec{k}) = \frac{1}{2}G_\mathrm{R}^{\mu\nu\rho\sigma}(\omega,\vec{k})\delta g_{\rho\sigma}(\omega,\vec{k}).   
\end{equation}

Furthermore, on the Minkowski background, the Ward identity \cref{eq:ward_id_general_coordinates} simplifies to 
\begin{equation}
\begin{split}
    & \frac{\partial}{\partial x^\mu} G_\mathrm{R}^{\mu\nu\rho\sigma}(x - y) + \left[ \bar T^{\mu\rho}(x) \eta^{\nu\sigma} + \bar T^{\mu\sigma}(x) \eta^{\nu\rho} \right. \\ & \left. - \bar T^{\rho\sigma}(x) \eta^{\nu\mu} \right] \frac{\partial}{\partial x^\mu} \delta^{(d)}(x-y) = 0.
\end{split}
\end{equation}
In terms of the Fourier space representation, with $k^\mu~=~(\omega, \mathbf{k})$ this becomes 
\begin{equation}
    k_\mu \left[G_\mathrm{R}^{\mu\nu\rho\sigma}(k) + \bar T^{\mu\rho}\eta^{\nu\sigma} + \bar T^{\mu\sigma}\eta^{\nu\rho} - \bar T^{\rho\sigma}\eta^{\mu\nu}    \right] = 0,
\end{equation}
in agreement with the findings in refs.~\cite{Czajka:2017,Jeon:2025} if one takes into account their chosen conventions for the metric signature and linear response equation.

%% file: inputs/hydro.tex
\subsection{Equations of motion in Mueller-Israel-Stewart theory}\label{sec:IsraelStewart}

We start by briefly recalling the equations of motion of relativistic fluid dynamics.
The degrees of freedom are taken to be the ones of the symmetric energy-momentum tensor, that is decomposed in the Landau frame~\cite{LANDAU} as
\begin{equation}
    T^{\mu\nu} = \epsilon u^\mu u^\nu + \left( p+\pi_\text{bulk} \right)\Delta^{\mu\nu} + \pi^{\mu\nu}.
    \label{eq:eom}
\end{equation}
Here, the fluid velocity $u^\mu$ is defined to be the time-like eigenvector, $T^\mu_{\phantom{\mu}\nu} u^\nu = - \epsilon u^\mu$, where the eigenvalue defines the (internal) energy density $\epsilon$. 
We normalize the fluid velocity to $g_{\mu\nu} u^\mu u^\nu = -1$. 
We are also using the projector orthogonal to the fluid velocity $\Delta^{\mu\nu} = u^\mu u^\nu + g^{\mu\nu}$.

In the decomposition in \cref{eq:eom}, $p$ is the equilibrium part of the pressure related to the energy density by the thermodynamic equilibrium equation of state $p(\epsilon)$. 
The remaining part on the diagonal is the bulk viscous pressure $\pi_\text{bulk}$, and $\pi^{\mu\nu}$ is the shear stress that is symmetric, traceless, $g_{\mu\nu}\pi^{\mu\nu}=0$, and orthogonal to the fluid velocity $u_\mu \pi^{\mu\nu} = 0$. 

Four equations of motion for the ten degrees of freedom of the energy-momentum tensor follow from the covariant conservation law \eqref{eq:covariant_conservation_tmunu}. 
One can obtain the evolution equations for the energy density $\epsilon$, 
\begin{equation}
    u^\mu \partial_\mu \epsilon + (\epsilon + p + \pi_\text{bulk}) \nabla_\mu u^\mu + \pi^{\mu\nu} \nabla_\mu u_\nu  = 0,
\label{eq:eomEnergy}
\end{equation}
and the fluid velocity $u^\mu$,
\begin{equation}
    \begin{split}
        (\epsilon + p + \pi_\text{bulk}) u^\lambda \nabla_\lambda u^\mu + &\Delta^{\mu\nu} \partial_\nu (p+\pi_\text{bulk}) \\
        &+ \Delta^\mu_{\phantom{\mu}\nu} \nabla_\rho \pi^{\nu\rho}= 0.
    \end{split}
\label{eq:eomVelocity}
\end{equation}
These need to be supplemented by constitutive relations for the bulk viscous pressure $\pi_\text{bulk}$ and shear stress $\pi^{\mu\nu}$. 
For our present purpose it will be sufficient to take these to be of the form 
\begin{equation}
\begin{split}
    \tau_\text{bulk} u^\lambda \partial_\lambda \pi_\text{bulk} + \pi_\text{bulk} &= -\zeta \nabla_\rho u^\rho, \\
    \tau_\text{shear} \Delta^{\mu\nu}_{\phantom{\mu\nu}\alpha\beta}  u^\lambda \nabla_\lambda \pi^{\alpha\beta} + \pi^{\mu\nu} &= -2\eta \sigma^{\mu\nu},
\end{split}
\label{eq:constitutive}
\end{equation}
with the bulk viscosity $\zeta$, shear viscosity $\eta$, the relaxation times $\tau_\text{bulk}$ and $\tau_\text{shear}$ and the projector to symmetric and traceless tensors orthogonal to the fluid velocity $\Delta^{\mu\nu}_{\phantom{\mu\nu}\alpha\beta} = (1/2) \Delta^\mu_{\phantom{\mu}\alpha} \Delta^\nu_{\phantom{\mu}\beta} + (1/2) \Delta^\mu_{\phantom{\mu}\beta} \Delta^\nu_{\phantom{\mu}\alpha} - (1/(d-1)) \Delta^{\mu\nu} \Delta_{\alpha\beta}$. 
Finally, 
\begin{equation}
    \sigma^{\mu\nu} = \Delta^{\mu\nu\alpha}_{\phantom{\mu\nu\alpha}\beta} \nabla_\alpha u^\beta, 
\end{equation}
is a symmetric and traceless combination of covariant derivatives of the fluid velocity.

The shear stress $\pi^{\mu\nu}$ has five independent components, so that \eqref{eq:constitutive} provides the remaining six equations of motion.

\subsection{Ideal fluids on curved spacetimes}
Before turning to relativistic fluids with dissipation we may start with ideal fluids. 
Their response to metric perturbations is fully governed by thermodynamic properties like susceptibilities, of order zero in terms of a fluid dynamic gradient expansion.

From a quantum field theoretic point of view, thermal equilibrium states are characterized by a vector field $\beta^\mu(x)= u^\mu(x) / T(x)$, the ratio of fluid velocity and temperature~\cite{Calzetta:2008iqa}. 
More specifically, in the Matsubara formalism~\cite{Matsubara:1955ws}, an equilibrium state on a Cauchy hypersurface $\Sigma$ can be described in terms of coordinates $x^\mu - \iu \sigma \beta^\mu(x)$, where $\sigma$ with $0\leq \sigma<1$ is a dimensionless imaginary time coordinate, and bosonic (fermionic) fields are (anti-~) periodic in the sense that $\Phi(x-\iu\beta(x)) = \pm \Phi(x)$. 
The manifold formed by the Cauchy hypersurface and the imaginary time coordinate has a Euclidean metric $g^\text{E}_{\mu\nu}(x) = g_{\mu\nu}(x) + 2 u_\mu(x) u_\nu(x)$. 
For global thermal equilibrium states the vector field $\beta^\mu(x)$ needs to be a Killing vector field such that  
\begin{equation}
\begin{split}
  \mathcal{L}_\beta g_{\mu\nu} = & \beta^\rho \partial_\rho g_{\mu\nu} + g_{\rho\nu}\partial_\mu \beta^\rho + g_{\mu\rho} \partial_{\nu} \beta^\rho \\
  = & \nabla_\mu \beta_\nu + \nabla_\nu \beta_\mu = 0,
\end{split}
\end{equation}
where $\mathcal{L}_\beta$ is the Lie derivative in the direction of $\beta^\rho$.

Local thermal equilibrium states as they are used in an ideal fluid approximation can now be defined by releasing this restriction, i.e.\ we allow generic local fields $\beta^\mu(x)$ to define a state, but assume implicitly that it varies slowly in space and time such that the concepts of thermodynamics can be locally applied. 
We are then asking how the expectation value of the energy-momentum tensor $T^{\mu\nu}(x)$ reacts to variations of the spacetime metric $g_{\mu\nu}(x)$, under the assumption that $\beta^\mu(x)$, which defines a state through the above periodicity condition, remains fixed.

The local temperature is related to the vector field $\beta^\mu$ through the relation
\begin{equation}
    T = \frac{1}{\sqrt{-g_{\mu\nu}\beta^\mu\beta^\nu}}.
\end{equation}
The variation of the temperature at constant $\beta^\mu$ is accordingly 
\begin{equation}
    \delta T = \frac{1}{2}T^3\beta^\mu\beta^\nu \delta g_{\mu\nu}.
\label{eq:DeltaTVariationMetric}
\end{equation}

Energy density and pressure have the variations
\begin{align} 
    \delta\epsilon &= \frac{\partial \epsilon }{\partial T } \delta T = c_V\delta T, \\
    \delta p &= \frac{\partial p }{\partial T }\delta T =  s\delta T,
\end{align} 
where %we have used the thermodynamic relations $\left( \tfrac{\partial \epsilon }{\partial T } \right)_V = c_V$ and $\left( \tfrac{\partial p }{\partial T } \right) = s(T)$ for 
$c_V$ is the heat capacity density at constant volume, and $s$ is the entropy density.

For fixed $\beta^\mu$ one has the variation of the fluid velocity
\begin{equation}
  \delta u^\mu = \beta^\mu \delta T = \frac{1}{2} u^\mu u^\rho u^\sigma \delta g_{\rho\sigma},    
\end{equation}
where \cref{eq:DeltaTVariationMetric} was used in the second step.

The energy-momentum tensor of an ideal fluid can be written with $\epsilon+ p = s T$ as 
\begin{equation}
        T^{\mu\nu} = (\epsilon+p) u^\mu u^\nu + p g^{\mu\nu} =  s T^3 \beta^\mu \beta^\nu + p g^{\mu\nu}.
    \label{eq:tmunu_equilibrium_general_coordinates}
\end{equation}
Variation at fixed $\beta^\mu$ yields
\begin{equation}
    \begin{split}
        \delta T^{\mu\nu} =& \left[(c_V + 3 s) T^2 \beta^\mu \beta^\nu + sg^{\mu\nu}\right]\delta T + p \delta g^{\mu\nu} \\
        = & \frac{1}{2}{\Big[} - p \left(g^{\mu\rho}g^{\nu\sigma} + g^{\mu\sigma}g^{\nu\rho} \right) +   sT g^{\mu\nu} u^\rho u^\sigma \\
        & + (c_V+ 3 s) T u^\mu u^\nu u^\rho u^\sigma {\Big]}\delta g_{\rho\sigma}.
    \end{split}
    \label{eq:tmunu_order1_metric}
\end{equation}
We used here $\delta g^{\mu\nu} = - (1/2)(g^{\mu\rho} g^{\nu\sigma} + g^{\mu\sigma} g^{\nu\rho}) \delta g_{\rho\sigma}$. Using also $\delta\sqrt{g} = (1/2) \sqrt{g} g^{\rho\sigma}\delta g_{\rho\sigma}$ leads to
\begin{equation}
    \delta \mathfrak{T}^{\mu\nu} = \frac{1}{2}\sqrt{g} \chi^{\mu\nu\rho\sigma} \delta g_{\rho\sigma},
\end{equation}
with the static energy-momentum susceptibility tensor
\begin{equation}
    \begin{split}
        \chi^{\mu\nu\rho\sigma} = & p \left(g^{\mu\nu}g^{\rho\sigma} - g^{\mu\rho}g^{\nu\sigma} - g^{\mu\sigma}g^{\nu\rho}\right) \\
        &+ s T (u^\mu u^\nu g^{\rho\sigma} + g^{\mu\nu} u^\rho u^\sigma)\\
        &+ (c_V T + 3 sT) u^\mu u^\nu u^\rho u^\sigma.
    \end{split}
    \label{eq:response_order_zero} 
\end{equation}
Note that the latter is symmetric with respect to the interchange of index pairs $\mu\nu$ and $\rho\sigma$, i.e.\ $\chi^{\mu\nu\rho\sigma} = \chi^{\rho\sigma\mu\nu}$. 
This is due to the fact that the expectation value of the energy-momentum tensor can itself be obtained by a variation of a generating functional with respect to the metric.

We also note that the static susceptibility tensor in \cref{eq:response_order_zero} contains no spacetime derivatives. 
This will change when we go beyond the ideal fluid limit below.

\subsection{Linearized fluid equations of motion}
We provide now linearized equations of motion for the general case of a dissipative fluid as discussed in section \ref{sec:IsraelStewart}. 
These linearized equations of motion determine to a large extend the response function defined in \cref{eq:response_function_definition}. 

At linear order, and using an equation of state $p=p(\epsilon)$, the fluid variables can be written as 
\begin{equation}
\begin{split}
    \epsilon =& \; \bar{\epsilon} + \delta \epsilon,\\
    p =& \; \bar{p} + \smash{c_\mathrm{S}^2} \delta \epsilon, \\
    u^\mu = & \; \bar u^\mu + \delta u^\mu, \\
    \pi_\text{bulk} = & \; \bar \pi_\text{bulk} + \delta \pi_\text{bulk}, \\
    \pi^{\mu\nu} = & \; \bar \pi^{\mu\nu}  + \delta \pi^{\mu\nu},
\end{split}
\label{eq:fluidFieldBackgroundFluctDecomposition}
\end{equation}
with the speed of sound defined by $c_\mathrm{S}^2=\partial \bar p /\partial \bar \epsilon$. 
We note that $u^\mu u^\nu g_{\mu\nu} = \bar u^\mu \bar u^\nu \bar g_{\mu\nu} = -1$ implies to linear order in perturbations $2\bar g_{\mu\nu} \bar u^\mu \delta u^\nu = - \bar u^\mu \bar u^\nu \delta g_{\mu\nu}$, and $u_\mu\pi^{\mu\nu} = 0$ implies $\bar u_\mu \delta \pi^{\mu\nu} = 0$. 

From the equation of motion for the energy density \eqref{eq:eomEnergy} we obtain
\begin{equation}
    \begin{split}
     &u^\mu\partial_\mu \delta \epsilon + \delta u^\mu \partial_\mu \bar \epsilon  \\
     +& (\bar\epsilon + \bar p + \bar\pi_\text{bulk})\left[ \nabla_\mu \delta u^\mu + \delta \Gamma_{\mu\lambda}^\mu \bar u^\lambda \right] \\
     +& \left[(1 + \smash{c_\mathrm{S}^2})\delta \epsilon + \delta \pi_\text{bulk}\right]\nabla_\mu \bar u^\mu \\
     +&\delta \pi^{\mu\nu}\nabla_\mu \bar u_\nu + \bar \pi^{\mu\nu} \left[ \nabla_\mu \delta u_\nu  - \delta\Gamma_{\mu\nu}^\lambda \bar u_\lambda \right] = 0,
    \end{split}
    \label{eq:energy_linearized_general_coords}
\end{equation}
while for the fluid velocity we find from \cref{eq:eomVelocity}
\begin{equation}
    \begin{split}
        &(\bar\epsilon + \bar p + \bar \pi_\text{bulk}) \left[\bar u^\lambda \nabla_\lambda \delta u^\mu + \bar u^\lambda \delta \Gamma_{\lambda\sigma}^\mu \bar u^\sigma + \delta u^\lambda \nabla_\lambda \bar u^\mu \right]  \\
        &+\left[(1 + \smash{c_\mathrm{S}^2})\delta \epsilon + \delta \pi_\text{bulk}\right] \bar u^\lambda \nabla_\lambda \bar u^\mu \\
        &+\delta \Delta^{\mu\nu}\partial_\nu(\bar p + \bar \pi_\text{bullk}) + \bar \Delta^{\mu\nu}\partial_\nu\left( c_\mathrm{S}^2\delta\epsilon + \delta \pi_\text{bulk} \right)\\
        &+\delta \Delta^\mu_{\phantom{\mu}\nu} \nabla_\rho \bar\pi^{\nu\rho} + \Delta^\mu_{\phantom{\mu}\nu}\big[\partial_\rho\delta \pi^{\nu\rho} + \bar\Gamma_{\rho\lambda}^\nu \delta\pi^{\lambda\rho} \\
        &+\bar\Gamma_{\rho\lambda}^\rho\delta\pi^{\nu\lambda} + \delta\Gamma_{\rho\lambda}^\nu\bar\pi^{\lambda\rho} + \delta\Gamma_{\rho\lambda}^\rho\bar\pi^{\nu\lambda} \big] = 0  .
    \end{split}
    \label{eq:velocity_linearized_general_coords}
\end{equation}
Linearizing the constitutive equations \eqref{eq:constitutive} leads us to
\begin{equation}
    \begin{split}
    &\left[ 1+\tau_\text{bulk} \bar u^\lambda \partial_\lambda  \right]\delta \pi_\text{bulk} =\\
    &- \left[ \tau_\text{bulk} \delta u^\lambda \partial_\lambda + \frac{\partial \tau_\text{bulk}}{\partial T} \delta T \bar u^\lambda \partial_\lambda  \right]\bar \pi_\text{bulk}  \\
    &-\zeta\left( \nabla_\mu \delta u^\mu + \delta \Gamma_{\mu\lambda}^\lambda \bar u^\mu \right) + \frac{\partial\zeta}{\partial T} \delta T \nabla_\mu \bar u^\mu,
    \end{split}
    \label{eq:bulk_pressure_linearized_general_coords}
\end{equation}
and 
\begin{equation}
    \begin{alignedat}{2} 
    &\delta \pi^{\mu\nu} + \tau_\text{shear} \bar\Delta^{\mu\nu}_{\phantom{\mu\nu}\alpha\beta} \bar u^\lambda \nabla_\lambda \delta \pi^{\alpha\beta} \\
    &+ \left[ \frac{\partial\tau_\text{shear}}{\partial T} \delta T \bar \Delta^{\mu\nu}_{\phantom{\mu\nu}\alpha\eta} \bar u^\lambda \nabla_\lambda + \tau_\text{shear} \delta(\Delta^{\mu\nu}_{\phantom{\mu\nu}\alpha\beta})\bar u^\lambda \nabla_\lambda\right. \\
    &+\left.\tau_\text{shear} \bar \Delta^{\mu\nu}_{\phantom{\mu\nu}\alpha\beta}\delta u^\lambda \nabla_\lambda \right] \bar \pi^{\alpha\beta} + 2\tau_\text{shear}\bar\Delta^{\mu\nu}_{\phantom{\mu\nu}\alpha\beta} \bar u^\lambda \delta \Gamma_{\lambda\sigma}^\alpha \bar \pi^{\sigma\beta}  \\
    &=- 2 \frac{\partial\eta}{\partial T} \delta T \bar \Delta^{\mu\nu\phantom{\alpha}\beta}_{\phantom{\mu\nu}\alpha} \nabla_\beta \bar u^\alpha - 2\eta \left[ \delta(\Delta^{\mu\nu\phantom{\alpha}\beta}_{\phantom{\mu\nu}\alpha }) \nabla_\beta \bar u^\alpha \right.\\
    &+ \left. \bar \Delta^{\mu\nu\phantom{\alpha}\beta }_{\phantom{\mu\nu}\alpha} \delta \Gamma_{\beta\lambda}^\alpha \bar u^\lambda + \bar \Delta^{\mu\nu\phantom{\alpha}\beta}_{\phantom{\mu\nu}\alpha}\nabla_\beta \delta u^\alpha \right]. 
    \end{alignedat}
    \label{eq:shear_linearized_general_coords}
\end{equation}
Note that all transport coefficients do have a spacetime dependence by their dependence on the temperature.

Using \cref{eq:bulk_pressure_linearized_general_coords,eq:shear_linearized_general_coords,eq:energy_linearized_general_coords,eq:velocity_linearized_general_coords}, it is possible to determine the response function of the system or directly extract Kubo formulas. 
The latter has already been done for Mueller-Israel-Stewart theories on curved spacetimes in refs.~\cite{Moore:2010bu,Jeon:2025}. 

%% file: inputs/calculations.tex
We will now address the inversion problem to determine the response function from the linearized equations of motion for the fluid fields, \cref{eq:bulk_pressure_linearized_general_coords,eq:shear_linearized_general_coords,eq:energy_linearized_general_coords,eq:velocity_linearized_general_coords}.

The calculation will be done on a Minkowski spacetime background $\bar g_{\mu\nu}(x) = \eta_{\mu\nu}$, filled by a homogeneous fluid in global thermal equilibrium with the energy-momentum tensor given in \cref{eq:ideal_fluid}. 
Consequently, the background Christoffel symbols $\bar \Gamma_{\mu\nu}^\alpha$ vanish, as well as all dissipative fluid fields on the background level, $\bar{\pi}_\text{bulk}=\bar{\pi}^{\mu\nu}=0$. 
Furthermore, we work in the frame where $\bar u^\mu = (1,\vec{0})$.

\subsection{Projections from energy-momentum tensor density to fluid fields}
We now work out how the perturbations in the fluid fields defined through the splitting in \cref{eq:fluidFieldBackgroundFluctDecomposition} are related to the perturbation in the energy-momentum tensor density defined by the splitting in \cref{eq:SplittingTFrak} and \eqref{eq:defFrakT}. 
These relations are important ingredients for a determination of the response function in \cref{eq:response_function_definition} from the (linearized) fluid dynamic equations of motion.

The perturbation in the energy-momentum tensor density is on a Minkowski space background related to the perturbation of the energy-momentum tensor through
\begin{equation}
    \delta \mathfrak{T}^{\mu\nu} = \frac{1}{2}\eta^{\rho\sigma}\delta g_{\rho\sigma} \bar T^{\mu\nu} + \delta T^{\mu\nu},
\end{equation}
where the latter can in turn be written as  
\begin{equation}
    \begin{split}
    \delta T^{\mu\nu} = & \left[ \bar u^\mu \bar u^\mu + \smash{c_\mathrm{S}^2}\bar \Delta^{\mu\nu} \right]\delta \epsilon \\
    &+ (\bar \epsilon + \bar p )\left[ \bar u^\mu \delta u^\nu + \bar u^\nu \delta u^\mu  \right] \\
    & + \delta \pi^{\mu\nu} + \bar \Delta^{\mu\nu} \delta \pi_\text{bulk} + \bar p \delta g^{\mu\nu}.
    \end{split}
\end{equation}
From this we can read off the projection prescriptions to access the fluid dynamical perturbation fields $\delta\epsilon, \delta u^i, \delta \pi^{ij}$ and $\delta\pi_\text{bulk}$ from the perturbations $\delta \mathfrak{T}^{\mu\nu}$ and metric perturbations, 
\begin{equation}
    \begin{split}
        \delta \epsilon =& \delta \mathfrak{T}^{00} - \bar p \delta g^{00} - \bar \epsilon \delta \sqrt{g}, \\
        \delta u^i =& \frac{1}{\bar \epsilon + \bar p } \delta \mathfrak{T}^{0i}  - \frac{\bar p }{\bar \epsilon + \bar p }\delta g^{0i}, \\
        \delta \pi_\text{bulk} =& \frac{1}{d-1}\delta_{ij}\delta\mathfrak{T}^{ij} - c_\mathrm{S}^2 \delta \mathfrak{T}^{00} - \frac{\bar p }{d-1}\delta_{ij} \delta g^{ij}\\
        &+ c_\mathrm{S}^2 \bar p \delta g^{00} + (c_\mathrm{S}^2\bar\epsilon - \bar p )\delta \sqrt{g},\\
        \delta \pi^{ij} =& \Delta^{ij}_{\phantom{ij}mn}\delta \mathfrak{T}^{mn} - \bar p \Delta^{ij}_{\phantom{ij}mn}\delta g^{mn}.
    \end{split}
    \label{eq:projectors}
\end{equation}
For the calculation of the response function we will use the linearized constitutive relations given by \cref{eq:bulk_pressure_linearized_general_coords,eq:shear_linearized_general_coords} together with the projectors \cref{eq:projectors} to write them as a system of equations of the spatial modes $\mathfrak{T}^{ij}$. 
Inverting this system of equations will lead the response of the spatial-spatial components for the energy-momentum tensor density $\mathfrak{T}^{ij}$ due to a general deviation $\delta g_{\alpha\beta}(x)$ of the metric.

Additionally, the conservation law \cref{eq:conservation_first_order_curved_spacetime} relates the temporal-spatial and temporal-temporal components of the energy-momentum tensor density to the spatial-spatial components via the Fourier space relations 
\begin{equation}
\begin{split} 
    \delta \mathfrak{T}^{00}(\omega, \mathbf{k}) =& \frac{k_i k_j}{\omega^2 }\delta \mathfrak{T}^{ij}(\omega, \mathbf{k}) + \frac{\bar T^{\mu\rho}}{\iu\omega }\delta \Gamma_{\mu\rho}^0(\omega, \mathbf{k})\\
    & + \frac{k_i }{\iu\omega^2 }\bar T^{\mu\rho} \delta \Gamma_{\mu\rho}^i(\omega, \mathbf{k}), 
\end{split}
\label{eq:eomRelations_momentum}
\end{equation}
and
\begin{equation}
\begin{split}
    \delta \mathfrak{T}^{0j}(\omega, \mathbf{k}) =& \frac{k_i }{\omega }\delta\mathfrak{T}^{ij}(\omega, \mathbf{k}) + \frac{\bar T^{\mu\rho}}{\iu\omega }\delta \Gamma_{\mu\rho}^j(\omega, \mathbf{k}).  
\end{split}
\label{eq:eomRelations_density}
\end{equation}
Using these relations, it suffices to calculate the response of the spatial-spatial components $\delta \mathfrak{T}^{ij}$, while the remaining components are related to those by
\begin{equation}
    \begin{split}
        2\delta \mathfrak{T}^{i0}(\omega, \mathbf{k}) =& G^{i0\alpha\beta}(\omega, \mathbf{k}) \delta g_{\alpha\beta}(\omega, \mathbf{k}),\\ 
        =& \frac{k_j }{\omega} G^{ij\alpha\beta}(\omega, \mathbf{k})\delta g_{\alpha\beta}(\omega, \mathbf{k})\\
        &+ \frac{2}{\iu\omega }\delta \Gamma_{\mu\rho}^i(\omega, \mathbf{k}) \bar T^{\mu\rho}, \label{eq:relation_t0i_to_response} \\
    \end{split}
\end{equation}
as well as
\begin{equation}
\begin{split}
    2\delta \mathfrak{T}^{00}(\omega, \mathbf{k}) =& G^{00\alpha\beta}(\omega, \mathbf{k}) \delta g_{\alpha\beta}(\omega, \mathbf{k})\\
    % =& \frac{k_i}{\omega^2}G^{i0\alpha\beta}(\omega, \mathbf{k})\delta g_{\alpha\beta}(\omega, \mathbf{k}) \\
    %  &+ \frac{2 \bar T^{\mu\rho} k_i }{\iu\omega^2 }\delta \Gamma_{\mu\rho}^i(\omega, \mathbf{k}) \\
    =& \frac{k_i k_j }{\omega^2 } G^{ij\alpha\beta}(\omega,\vec{k})\delta g_{\alpha\beta}(\omega,\vec{k}), \\
    &+ \frac{2\bar T^{\mu\rho} k_i }{\iu\omega^2 }\delta \Gamma_{\mu\rho}^i(\omega,\vec{k}) \\ 
    &+ \frac{2\bar T^{\mu\rho}}{\iu \omega }\delta \Gamma_{\mu\rho}^0(\omega,\vec{k}).
\end{split}\label{eq:relation_t00_to_response}
\end{equation}
Using the explicit expression for $\delta \Gamma_{\mu\rho}^\nu$, and $\bar T^{\mu\nu}$ given by \cref{eq:delta_christoffel,eq:tmunu_equilibrium_general_coordinates} respectively, one can check that the Ward identity \cref{eq:ward_id_general_coordinates} is indeed satisfied.

\subsection{Decomposition into orthogonal subspaces}
To invert the equations of motion given by \cref{eq:bulkpressure_linearized_tmn,eq:shearstress_linearized_tmn} we will decompose the spatial components of the energy-momentum tensor into parts which transform as scalars, vectors or tensors of rank two with respect to rotations around a reference momentum.
This reference momentum is defined by the unit vector $\smash{\hat k_i = |\vec{k}|^{-1} k_i}$ and the orthogonal projector $\smash{\hat \Delta_{ij} = \delta_{ij} - \hat k_i \hat k_j}$ such that $\smash{\hat\Delta_{ij} \hat k^j = 0}$.
We can decompose any symmetric tensor of rank two as 
\begin{equation}
    T^{ij} = t_1 \hat\Delta^{ij} + \hat t^{ij} + t_3 \hat k^i \hat k^j + \left( \hat v^m \hat k^n + \hat v^n \hat k^m  \right).
    \label{eq:tensor_decomp}
\end{equation}
Here $\hat t^{ij}$ are the traceless symmetric parts perpendicular to $\smash{\hat k_i}$ with $\smash{\hat t^{ij} = \hat t^{ji}}$ and $\smash{\hat t^{ij} \hat \Delta_{ij} = \hat t^{ij}\hat k_j = 0}$ while $v^l$ is a vector perpendicular to the momentum, $\smash{v^l \hat k_l = 0}$. 
The coefficients of the first and third term in \cref{eq:tensor_decomp} transform as scalars under rotations around the reference direction $\smash{\hat k_i}$, while the $\hat v^m$ coefficient of the fourth term transforms as a $\smash{(d-2)}$-component vector and the second term transforms as a $\smash{(d-2)\times(d-2)}$ matrix under rotations.

\subsection{Decomposition of the bulk pressure and shear stress equations of motion}
Starting with the constitutive relation of the bulk pressure \eqref{eq:bulk_pressure_linearized_general_coords} and switching to momentum space while expanding $\pi_\text{bulk}$, $u^\mu$ and $\nabla_\mu$ up to linear order around the background $\pi_\text{bulk}=0$, we find the relation
\begin{equation}
    \begin{split}
        (1-\iu\omega\tau_\text{bulk})\delta\pi_\text{bulk}(\omega,\vec{k}) =& -\zeta \iu k_\mu \delta u^\mu(\omega,\vec{k}) \\
        &- \zeta \delta\Gamma_{\mu \lambda}^\mu(\omega,\vec{k}) \bar u^\lambda,
    \end{split}
\end{equation}
where $\bar u^\mu = \delta_0^\mu$.
Utilizing the projections defined in \cref{eq:projectors} together with \cref{eq:eomRelations_density,eq:eomRelations_momentum}, we express the bulk pressure and fluid velocity as a function of $\delta\mathfrak{T}^{mn}$.
This results in the formal equation $C_{mn}(\omega,\vec{k})\delta\mathfrak{T}^{mn}(\omega,\vec{k}) = A^{\alpha\beta}(\omega,\vec{k})\delta g_{\alpha\beta}(\omega,\vec{k})$ such that 
\begin{equation}
    \begin{alignedat}{2}
        &\left[ c_1 \hat\Delta_{mn} + c_2 \hat k_m \hat k_n  \right]\delta \mathfrak{T}^{mn}(\omega,\vec{k}) = \\
        &\big[ a_1 u^\alpha u^\beta + a_2\hat \Delta^{\alpha\beta} + a_3\hat k^\alpha \hat k^\beta \\
        &+ a_4 \left( \hat k^\alpha u^\beta + u^\alpha \hat k^\beta \right) \big]\delta g_{\alpha\beta}(\omega,\vec{k}),
    \end{alignedat}
    \label{eq:bulkpressure_linearized_tmn}
\end{equation}
with the coefficients
\begin{equation}
    \begin{split}
        c_1 =& \frac{1-\iu\omega\tau_\text{bulk}}{d-1},\\
        c_2 =& \left( 1-\iu\omega\tau_\text{bulk} \right)\left( \frac{1}{d-1 } - c_\mathrm{S}^2\frac{\vec{k}^2 }{\omega^2 } \right) + \frac{\iu\zeta \vec{k}^2 }{\omega (\bar \epsilon + \bar p )},
    \end{split}
    \label{eq:coefficients_c}
\end{equation}
and
\begin{equation}
    \begin{split}
        a_1 =& -(1-\iu\omega\tau_\text{bulk})\left[\frac{\bar p }{2} + \frac{c_\mathrm{S}^2\bar \epsilon \vec{k}^2 }{2\omega^2 }  \right] + \frac{\iu\zeta\bar\epsilon \vec{k}^2 }{2\omega(\bar\epsilon + \bar p )  },\\
        a_2 =& (1-\iu\omega\tau_\text{bulk})\left[\frac{d-3 }{2(d-1) }\bar p -\frac{c_\mathrm{S}^2 }{2}(\bar \epsilon + \bar p )\right. \\
        &-\left. \frac{c_\mathrm{S}^2 \bar p \vec{k}^2 }{2\omega^2 }\right] + \frac{1}{2}\iu\omega\zeta + \frac{\iu\zeta\bar p \vec{k}^2 }{2\omega(\bar\epsilon + \bar p ) },\\
        a_3 =& \left( 1-\iu\omega\tau_\text{bulk} \right)\left[ \frac{d-3 }{2(d-1 )}\bar p - \frac{c_\mathrm{S}^2 }{2}(\bar \epsilon + \bar p )\right. \\ 
        &+ \left.\frac{c_\mathrm{S}^2\bar p \vec{k}^2 }{2\omega ^2 } \right]+\frac{\iu\omega\zeta }{2} - \frac{\iu\zeta\bar p \vec{k}^2 }{2\omega(\bar \epsilon + \bar p )} ,\\
        a_4 =& \frac{\iu\zeta |\vec{k}|}{2 } - (1-\iu\omega\tau_{\text{bulk}})c_\mathrm{S}^2(\bar \epsilon + \bar p )\frac{|\vec{k}|}{2\omega }.
    \end{split}
    \label{eq:coefficients_a}
\end{equation}

Similarly, using \cref{eq:shear_linearized_general_coords}, we find for the shear stress in momentum space
\begin{equation}
    \begin{split}
        (1-\iu\omega \tau_\text{shear})\delta \pi^{ij}(\omega,\vec{k}) =& -2\eta \Delta^{ijl}_{\phantom{ijl}m} \left[\iu k_l \delta u^m(\omega,\vec{k}) \right.\\
        &+ \left. \delta \Gamma_{l\lambda }^m(\omega,\vec{k}) \bar u^\lambda\right].
    \end{split}
\end{equation}
Again using \cref{eq:eomRelations_momentum,eq:projectors} we can write this as $\smash{D^{ij}_{\phantom{ij}mn}(\omega,\vec{k})\delta\mathfrak{T}^{mn}(\omega,\vec{k})=~B^{ij\alpha\beta}(\omega,\vec{k})\delta g_{\alpha\beta}(\omega,\vec{k})}$ with the tensor decomposition
\begin{equation}
    \begin{split}
        &\left[ d_1 \Delta^{ij}_{\phantom{ij}mn} + d_2 \hat k_l \left( \Delta^{ijl}_{\phantom{ijl}m}\hat k_n + \Delta^{ijl}_{\phantom{ijl}n}\hat k_m \right) \right]\delta\mathfrak{T}^{mn}(\omega,\vec{k}) = \\
        &\big[b_1 \Delta^{ij\alpha\beta} + b_2 \hat k_m \left( \Delta^{ijm\alpha} u^\beta + \Delta^{ijm\beta} u^\alpha \right) \\
        &+ b_3 \Delta^{ijmn}\hat k_m \hat k_n u^\alpha u^\beta \\
        &+ b_4 \hat k_m \left( \Delta^{ijm\alpha}\hat k^\beta + \Delta^{ijm\beta}\hat k^\alpha  \right) \\
        &+ b_5 \Delta^{ijmn}\hat k_m \hat k_n \hat k^\alpha \hat k^\beta \\
        &+ b_6 \Delta^{ijmn}\hat k_m \hat k_n \hat \Delta^{\alpha\beta} \big]\delta g_{\alpha\beta}(\omega,\vec{k}),
    \end{split}
    \label{eq:shearstress_linearized_tmn}
\end{equation}
and coefficients 
\begin{equation}
        d_1 = 1-\iu\omega\tau_\text{shear}, \qquad d_2 = \frac{\iu\eta \vec{k}^2}{\omega(\bar \epsilon + \bar p )},
    \label{eq:coefficients_d}
\end{equation}
as well as
\begin{equation}
    \begin{split}
        b_1 =& -\bar p  (1-\iu\omega\tau_\text{shear}) + \iu\eta\omega,\qquad b_2 = \iu\eta|\vec{k}|,\\
        b_3 =& \frac{\iu\eta\bar\epsilon \vec{k}^2}{\omega(\bar \epsilon + \bar p)}, \qquad b_4 = -\frac{\iu\eta\bar p \vec{k}^2}{\omega(\bar \epsilon + \bar p)},\\
        b_6 =& b_5 = -b_4.%\frac{\iu\eta\bar p }{\omega(\bar \epsilon + \bar p)}.
    \end{split}
    \label{eq:coefficients_b}
\end{equation}
To calculate the response of the spatial components of $\mathfrak{T}^{ij}$ to some external perturbation $\delta g_{\alpha\beta}$, both \cref{eq:bulkpressure_linearized_tmn,eq:shearstress_linearized_tmn} have to be inverted which results in $\smash{G_\mathrm{R}^{ij\alpha\beta}(\omega,\vec{k})}$.

The linear maps $A^{\alpha\beta}$, $B^{mn\alpha\beta}$, $C_{mn}$ and $D_{mn}^{ij}$ defined via \cref{eq:bulkpressure_linearized_tmn} and \cref{eq:shearstress_linearized_tmn} can be decomposed into maps between the specific orthogonal subspaces. 
In particular, we can split all maps into projections of vectors, symmetric-traceless matrices and scalars with respect to rotations around $\smash{\hat k^i}$.
The latter will mix the scalars contained in the trace of the tensor of rank two.

We can decompose the maps by defining projectors for specific subspaces with   
\begin{equation}
\begin{split}
    \mathcal{P}^{ij}_{\mathrm{(S1)}mn} =& \frac{1}{d-2}\hat\Delta^{ij}\hat\Delta_{mn}, \\
    \mathcal{P}^{ij}_{\mathrm{(S2)}mn} =& \hat k^i \hat k^j \hat k_m \hat k_n ,\\
    \mathcal{P}^{ij}_{\mathrm{(V)}mn} =& \frac{1}{2}\left(\hat k^i \hat\Delta^j_m \hat k_n + \hat k^j \hat \Delta^i_m \hat k_n \right. \\
    &\quad\,+ \left.\hat k^i \hat \Delta^j_n \hat k_m + \hat k^j \hat \Delta^i_n \hat k_m  \right), \\
    \mathcal{P}^{ij}_{\mathrm{(TT)}mn} =& \frac{1}{2}\left( \hat\Delta^i_m\hat\Delta^j_n + \hat\Delta^i_n\hat\Delta^j_m\right. \\
    &\quad\;- \left. \frac{2}{d-2}\hat \Delta^{ij}\hat \Delta_{mn} \right), \\
    \mathcal{P}^{ij}_{\mathrm{(S)}mn} =& \mathcal{P}^{ij}_{\mathrm{(S1)}mn} + \mathcal{P}^{ij}_{\mathrm{(S2)}mn},
\end{split}
\label{eq:projectors_orthogonal_subspaces}
\end{equation}
where $\smash{\mathcal{P}_{\mathrm{(S1)}}}$ is a map between matrices proportional to the orthogonal projector $\smash{\hat\Delta_{ij}}$, $\smash{\mathcal{P}_{\mathrm{(S2)}}}$ for the scalar in the $\hat k_i$ direction, $\smash{\mathcal{P}_{\mathrm{(V)}}}$ maps vectors onto vectors and $\smash{\mathcal{P}_{\mathrm{(TT)}}}$ between symmetric traceless tensors. 
The projector onto the scalar subspace is just given by the direct sum of the two scalar projectors (S1) and (S2).
% One may verify that indeed these objects define projectors onto the respective subspace and are each orthogonal to one another. 

Using the projectors on the equations of motion we find the four equations
\begin{align}
    D^{ij}_{\mathrm{(TT)}mn}\delta \mathfrak{T}^{mn}_{\mathrm{(TT)}} &= B^{ij\alpha\beta}_{\mathrm{(TT)}}\delta g_{\alpha\beta} \label{eq:eom_shearstress_linearized_tt},\\
    D^{ij}_{\mathrm{(V)}mn}\delta\mathfrak{T}^{mn}_{\mathrm{(V)}} &= B^{ij\alpha\beta}_{\mathrm{(V)}}\delta g_{\alpha\beta} \label{eq:eom_shearstress_linearized_v},\\
    D^{ij}_{\mathrm{(S)}mn}\delta\mathfrak{T}^{mn}_{\mathrm{(S)}} &= B^{ij\alpha\beta}_{\mathrm{(S)}}\delta g_{\alpha\beta} \label{eq:eom_shearstress_linearized_s},\\
    C_{(S)mn}\delta\mathfrak{T}^{mn}_{\mathrm{(S)}} &= A^{\alpha\beta}_{\mathrm{(S)}}\delta g_{\alpha\beta} \label{eq:eom_bulkpressure_linearized_s}.
\end{align}
Here we introduced the notation $\delta \mathfrak{T}_\mathrm{(L)}^{ij}=\mathcal{P}^{ij}_{\mathrm{(L)}mn}\delta \mathfrak{T}^{mn}$ for L referring to the S, V or TT label. 
Additionally, we define $\smash{D_{\mathrm{(TT)}mn}^{ij} = \mathcal{P}^{ijab}_{\mathrm{(TT)}} D_{abmn}}$ and $\smash{D_{\mathrm{(V)}mn}^{ij}=\mathcal{P}^{ijab}_{\mathrm{(V)}} D_{abmn}}$ while the scalars are given by the remainder $\smash{D_{\mathrm{(S)}mn}^{ij} = D_{\phantom{ij}mn}^{ij} - D_{\mathrm{(TT)}mn}^{ij} - D_{\mathrm{(V)}mn}^{ij}}$, and analogously define the parts $B_\mathrm{(TT)}$, $B_\mathrm{(V)}$ and $B_{\mathrm{(S)}}$. 
Furthermore, since the bulk pressure equation (\ref{eq:bulkpressure_linearized_tmn}) only contributes to the trace, we find that $\smash{C_{\mathrm{(S)}}=C}$ and $\smash{A_{\mathrm{(S)}}=A}$.
The explicit forms of the specific maps can be found in~\cref{sec:appendix_projectors}.

While \cref{eq:eom_shearstress_linearized_tt,eq:eom_shearstress_linearized_v} can be inverted independently on their respective subspaces, the scalars appear in both the shear stress and bulk pressure equations. 
Therefore, we need to invert the combination of both \cref{eq:eom_shearstress_linearized_s,eq:eom_bulkpressure_linearized_s}.

\subsection{Inversion of the equations of motion}
Starting with the orthogonal subspaces (TT) and (V) we can directly invert the shear modes since the bulk modes do not contribute to these spaces. 
Using the Fourier representation of \cref{eq:linear_response_definition} we find the response of the (TT) and (V) components of the energy-momentum tensor to be
\begin{equation}
    \begin{split}
    2\delta\mathfrak{T}^{ij}_{\mathrm{(TT)}}(\omega,\vec{k}) =& G_{\mathrm{(TT)}}^{ij\alpha\beta}\delta g_{\alpha\beta}(\omega,\vec{k})\\
    =& 2 \frac{1}{d_1 } B_{\mathrm{(V)}}^{ij\alpha\beta}(\omega,\vec{k})\delta g_{\alpha\beta}(\omega,\vec{k})\\
    =& 2\frac{b_1 }{d_1 }\mathcal{P}^{ij\alpha\beta}_{\mathrm{(TT)}}(\omega,\vec{k})\delta g_{\alpha\beta}(\omega,\vec{k}) \label{eq:response_tt},\\
    \end{split}
\end{equation}
and
\begin{equation}
    \begin{split}
    2\delta\mathfrak{T}^{ij}_{\mathrm{(V)}}(\omega,\vec{k}) =& G_{\mathrm{(V)}}^{ij\alpha\beta}(\omega,\vec{k})\delta g_{\alpha\beta}(\omega,\vec{k}) \\
    =& 2\frac{1}{d_1 + d_2 }B^{ij\alpha\beta}_{\mathrm{(V)}}(\omega,\vec{k})\delta g_{\alpha\beta}(\omega,\vec{k})\label{eq:response_v},\\
    =& 2\frac{b_1 + b_4 }{d_1 + d_2 }\mathcal{P}^{ij\alpha\beta}_{\mathrm{(V)}}(\omega,\vec{k})\delta g_{\alpha\beta}(\omega,\vec{k}) \\
    &+\frac{b_2 }{d_1 + d_2 }\Big( \hat k^{i} \hat \Delta^{j\alpha}u^{\beta} + \\
    &\hat k^{i} \hat \Delta^{j\beta}u^{\alpha} + \hat k^{j} \hat \Delta^{i\alpha}u^{\beta}\\
    &+ \hat k^{j} \hat \Delta^{i\beta}u^{\alpha}  \Big)\delta g_{\alpha\beta}(\omega,\vec{k}).
    \end{split}
\end{equation}

In order to invert the scalars we need to combine the shear and bulk pressure equations. 
Parameterizing the scalar parts of the energy-momentum tensor, using the scalar projector from \cref{eq:projectors_orthogonal_subspaces}, as 
\begin{equation}
    \begin{split}
        \delta\mathfrak{T}^{mn}_{\mathrm{(S)}} &= \hat k_i \hat k_j \delta\mathfrak{T}^{ij} \hat k^m \hat k^n + (d-2)^{-1}\hat{\Delta}_{ij}\delta\mathfrak{T}^{ij} \hat\Delta^{mn} \\
        &=\delta\mathfrak{T}^{mn}_{\mathrm{(S2)}}+\delta \mathfrak{T}^{mn}_{\mathrm{(S1)}},
    \end{split}
\end{equation} 
the bulk pressure \cref{eq:bulkpressure_linearized_tmn} yields the relation
\begin{equation}
    \begin{split}
        c_1 \hat{\Delta}_{ij}\delta\mathfrak{T}_{\mathrm{(S)}}^{ij}(\omega,\vec{k}) + c_2 \hat k_i \hat k_j \delta \mathfrak{T}_{\mathrm{(S)}}^{ij}(\omega,\vec{k})  = \\
        A^{\alpha\beta}(\omega,\vec{k})\delta g_{\alpha\beta}(\omega,\vec{k}).
    \end{split}
\end{equation}

The shear stress equations can be split into the two equations 
\begin{equation}
    \begin{split}
        \hat k_i \hat k_j D^{ij}_{(S)mn}(\omega,\vec{k})&\delta\mathfrak{T}^{mn}_{\mathrm{(S)}}(\omega,\vec{k}) = \\
        &\hat k_i \hat k_j B^{ij\alpha\beta}_{\mathrm{(S)}}(\omega,\vec{k})\delta g_{\alpha\beta}(\omega,\vec{k}),  \\
        \hat\Delta_{ij} D^{ij}_{(S)mn}(\omega,\vec{k})&\delta\mathfrak{T}^{mn}_{\mathrm{(S)}}(\omega,\vec{k}) = \\ 
        &\hat\Delta_{ij} B^{ij\alpha\beta}_{\mathrm{(S)}}(\omega,\vec{k})\delta g_{\alpha\beta}(\omega,\vec{k}) .
    \end{split}
\end{equation}
Plugging in the tensor decomposition of $D^{mn\alpha\beta}$ one can show that both equations are equivalent, and the scalar contributions are determined by the bulk pressure equation and only one of the shear stress equations.
Combining both leads to a linear system of equations,
\begin{equation}
    \begin{split}
        c_2 \hat k_i \hat k_j \delta\mathfrak{T}^{ij} + c_1 \hat{\Delta}_{ij}\delta\mathfrak{T}^{ij} =& A^{\alpha\beta}\delta g_{\alpha\beta},
    \end{split}
\end{equation}
\begin{equation}
    \begin{split}
        &\left[\left( 1-\frac{1}{d-1} \right)(d_1 + 2d_2)\hat k_i \hat k_j  -\frac{d_1 }{d - 1}\hat{\Delta}_{ij}\right]\delta\mathfrak{T}^{ij} \\
        &= \hat k_i \hat k_j B^{ij\alpha\beta}_{\mathrm{(S)}}\delta g_{\alpha\beta},  
    \end{split}
\end{equation}
which has the solution 
\begin{equation}
    \begin{alignedat}{2}
        \hat k_i \hat k_j \delta\mathfrak{T}^{ij}  &= \frac{1}{M} \bigg(&& c_1(d-1)\hat k_i \hat k_j B^{ij\alpha\beta}_{\mathrm{(S)}}\delta g_{\alpha\beta} \\
        &&&+  d_1 A^{\alpha\beta}\delta g_{\alpha\beta} \bigg),\\
        \hat{\Delta}_{ij}\delta\mathfrak{T}^{ij} &= \frac{1}{M}\bigg(&& (d-2)(d_1 + 2d_2 ) A^{\alpha\beta}\delta g_{\alpha\beta} \\
        &&&- c_2(d-1) \hat k_i \hat k_j B^{ij\alpha\beta}_{\mathrm{(S)}}\delta g_{\alpha\beta}\bigg),
    \end{alignedat}
    \label{eq:_solution_inversion_scalars}
\end{equation}
where $\mathrm{M} = c_2 d_1 + (d-2) c_1 (d_1 + 2 d_1)$.
The solution of both scalar parts (S1) and (S2) gives then rise to the scalar contributions of the response function where 
\begin{equation}
    \begin{split}
        2\delta \mathfrak{T}^{ij}_{\mathrm{(S)}}(\omega,\vec{k}) =& G_{\mathrm{(S)}}^{ij\alpha\beta}(\omega,\vec{k})\delta g_{\alpha\beta}(\omega,\vec{k}),
        % =& \left[ G_{\mathrm{(S1)}}^{ij\alpha\beta}(\omega,\vec{k}) + G_{\mathrm{(S2)}}^{ij\alpha\beta}(\omega,\vec{k}) \right]\delta g_{\alpha\beta}(\omega,\vec{k})
    \end{split}
\end{equation}
with
\begin{equation}
    \begin{split}
        G_\mathrm{(S)}^{ij\alpha\beta}(\omega,\vec{k}) =& s_1 \mathcal{P}_\mathrm{(S1)}^{ij\alpha\beta} + s_2 \mathcal{P}_\mathrm{(S2)}^{ij\alpha\beta} \\
        &+ \hat k^i \hat k^j \bigg[s_3 \hat\Delta^{\alpha\beta} + s_4 u^\alpha u^\beta \\
        &+ s_5 \left( \hat k^\alpha u^\beta + u^\alpha \hat k^\beta  \right)\bigg] \\
        &+ \hat \Delta^{ij} \bigg[ s_6 \hat k^\alpha \hat k^\beta + s_7 u^\alpha u^\beta \\
        &+ s_8 \left( \hat k^\alpha u^\beta + u^\alpha \hat k^\beta  \right) \bigg],
    \end{split}
\end{equation}
with coefficients, using the decompositions \cref{eq:lhs_shear_equation_tensor_decomposition,eq:rhs_shear_equation_tensor_decomposition}, 
\begin{equation}
    \begin{split}
        s_1 =& \frac{2 }{M }\left[ (d+2d_2)(d-2)a_2 + c_2 (b_1 + (d-2)b_4) \right], \\
        s_2 =& \frac{2}{M }\left[ d_1 a_3 + c_1 (d-2)(b_1 + b_4 ) \right], \\
        s_3 =& \frac{2}{M }\left[ d_1 a_2 - c_1\left( b_1 + (d-2)b_4  \right) \right], \\
        s_4 =& \frac{2 }{M }\left[ d_1 a_1 + c_1  b_3 (d-2) \right], \\
        s_5 =& \frac{2 }{M }\left[ d_1 a_4 + c_1  b_2 (d-2) \right], \\
        s_6 =& \frac{2 }{M }\left[ (d_1 + 2 d_2 )a_3 - c_2 (b_1 + b_4 ) \right], \\
        s_7 =& \frac{2 }{M }\left[ (d_1 + 2 d_2 )a_1 - c_2 b_3  \right], \\
        s_8 =& \frac{2 }{M }\left[ (d_1 + 2d_2 )a_4 -c_2 b_2  \right].
    \end{split}
\end{equation}
% and the $\beta_i$ coefficients are defined by $\smash{B^{ij\alpha\beta}_{\mathrm{(S)}}}$ and can be found in \cref{sec:appendix_projectors}.
\subsection{Response function}
Combining \cref{eq:response_tt,eq:response_v,eq:_solution_inversion_scalars} we find the full response function for the spatial modes for a general external metric perturbation. 
We will additionally formulate everything for arbitrary spatial momentum $\vec{k}$.  
The full response function is therefore 
\begin{equation}
    \begin{split}
            &G_\mathrm{R}^{ij\alpha\beta}(\omega,\vec{k}) = 2 \frac{b_1 }{d_1 }\mathcal{P}_{\mathrm{(TT)}}^{ij\alpha\beta} + 2\frac{b_1 + b_4 }{d_1 + d_2 }\mathcal{P}_{\mathrm{(V)}}^{ij\alpha\beta}\\
            &+ \frac{b_2 }{(d_1 + d_2 ) }\left( \hat k^i \hat\Delta^{j\alpha} u^\beta + \hat k^j \hat\Delta^{i\alpha}u^\beta + \hat k^i \hat\Delta^{j\beta}u^\alpha \right. \\
            &+ \left. \hat k^j\hat\Delta^{i\beta}u^\alpha \right) + s_1 \mathcal{P}_{\mathrm{(S1)}}^{ij\alpha\beta} + s_2 \mathcal{P}_{\mathrm{(S2)}}^{ij\alpha\beta} \\
            &+ \hat k^i \hat k^j \left[s_3 \hat \Delta^{\alpha\beta} + s_4 u^\alpha u^\beta + s_5\left( \hat k^\alpha u^\beta + u^\alpha \hat k^\beta  \right)\right] \\
            &+ \hat\Delta^{ij}\left[ s_6 \hat k^\alpha \hat k^\beta + s_7 u^\alpha u^\beta + s_8 \left( \hat k^\alpha u^\beta + u^\alpha \hat k^\beta  \right)  \right].
    \end{split}
    \label{eq:responsefunction_spatialspatial}
\end{equation}
Using \cref{eq:responsefunction_spatialspatial} together with the relations \cref{eq:relation_t0i_to_response,eq:relation_t00_to_response} we can construct the missing parts of the response function containing temporal-indices. 
The conservation law will introduce contact terms arising from the connection coefficient symbols $\delta \Gamma^\nu_{\mu\rho}$ that are demanded by the Ward identity. 
This leads to the expressions
\begin{equation}
    \begin{split}
        G_\mathrm{R}^{0j\alpha\beta}(\omega,\vec{k}) &= \frac{|\vec{k}|}{\omega }\bigg[ \frac{b_1 +  b_4 }{d_1 +  d_2 }\left( \hat\Delta^{j\alpha}\hat k^\beta + \hat\Delta^{j\beta }\hat k^\alpha  \right) \\
        &+\frac{b_2 }{d_1 +  d_2 }\left( \hat \Delta^{j\alpha} u^\beta + \hat\Delta^{j\beta}u^\alpha  \right) \\
        &+ \hat k^j \bigg( s_2 \hat k^\alpha \hat k^\beta + s_5 \left( \hat k^\alpha u^\beta + u^\alpha \hat k^\beta  \right) \\
        &+ s_3 \hat \Delta^{\alpha\beta} + s_4 u^\alpha u^\beta  \bigg)\bigg]\\
        &-\bar\epsilon\hat k^j \left( u^\alpha \hat k^\beta + \hat k^\alpha u^\beta  \right) - \bar\epsilon \frac{|\vec{k}| }{\omega }\hat k^j u^\alpha u^\beta \\
        &- \bar\epsilon \left( u^\alpha \hat \Delta^{\beta j } + u^\beta \hat \Delta^{\alpha j}  \right) + \frac{|\vec{k}| }{\omega }\bar p \bigg( \hat k^j \hat k^\alpha \hat k^\beta \\
        &+ \left( \hat k^\alpha \hat\Delta^{\beta j } + \hat k^\beta \hat \Delta^{\alpha j }  \right) - \hat k^j\hat \Delta^{\alpha\beta} \bigg),
    \end{split}
    \label{eq:responsefunction_timespatial}
\end{equation}
and 
\begin{equation}
    \begin{split}
        G_\mathrm{R}^{00\alpha\beta}(\omega,\vec{k}) =& \frac{\vec{k}^2}{\omega^2 }\bigg[s_2 \hat k^\alpha \hat k^\beta + s_3 \hat\Delta^{\alpha\beta} + s_4 u^\alpha u^\beta \\
        &+ s_5 \left( \hat k^\alpha u^\beta + \hat k^\beta u^\alpha  \right) \bigg] + \bar \epsilon u^\alpha u^\beta \\
        &-\bar\epsilon \frac{|\vec{k}| }{\omega }\left( \hat k^\alpha u^\beta + u^\alpha \hat k^\beta  \right) - \bar\epsilon \frac{\vec{k}^2 }{\omega^2 } u^\alpha u^\beta \\
        &- \frac{\bar p }{\omega} \bigg(|\vec{k}|\left( \hat k^\alpha u^\beta + u^\alpha \hat k^\beta  \right) \\
        &+ \omega \left( \hat \Delta^{\alpha\beta} + \hat k^\alpha \hat k^\beta  \right)\bigg) \\
        &+ \bar p \frac{\vec{k}^2 }{\omega^2 }\left( \hat k^\alpha \hat k^\beta - \hat \Delta^{\alpha\beta} \right).
    \end{split}
    \label{eq:responsefunction_timetime}
\end{equation}
The retarded correlator is completely determined by \cref{eq:responsefunction_spatialspatial,eq:responsefunction_timespatial,eq:responsefunction_timetime}. 
Using these equations one can now determine quantities like the static susceptibilities or Kubo formulas which relate the imaginary part of the response function to transport coefficients. 
The former is accessed by taking the static limit 
\begin{equation}
    \begin{split}
        \lim_{|\vec{k}|\rightarrow 0}G_\mathrm{R}^{\mu\nu\alpha\beta}(0,\vec{k})  =& p\left( \eta^{\mu\nu}\eta^{\alpha\beta} - \eta^{\mu\alpha}\eta^{\nu\beta} - \eta^{\mu\beta}\eta^{\nu\alpha}  \right)\\
        &+sT\left( \eta^{\mu\nu} u^\alpha u^\beta + u^\mu u^\nu \eta^{\alpha\beta} \right) \\
        &+(c_V T + 3sT)u^\mu u^\nu u^\alpha u^\beta,
    \end{split}
\end{equation}
coinciding with the Minkowski spacetime limit of \cref{eq:response_order_zero} and results obtained in~\cite{Czajka:2017,Jeon:2025}.
Kubo formulas for the shear or bulk viscosity can usually be found by expanding the response function at zero spatial momentum for small frequencies. 
This yields for example, taking the momentum in the 3-direction, 
\begin{equation}
    G^{1212}_\mathrm{R}(\omega,\vec{0}) \approx -\bar p + \iu\eta\omega  - \iu\eta\tau_\text{shear} \omega^2, \\
\end{equation}
and 
\begin{equation}    
    \begin{split}
        \frac{1}{d-1}\left[ G_\mathrm{R}^{3333}(\omega,\vec{0}) + (d-2)G_\mathrm{R}^{3322}(\omega,\vec{0}) \right] \approx \\
        \frac{d-3 }{d-1 }\bar p - c_\mathrm{S}^2(\bar\epsilon + \bar p ) + \iu\omega\zeta - \iu\zeta\tau_\text{bulk}\omega^2,
    \end{split}
    \label{eq:kubo_formula_zeta_kovtun}
\end{equation}
where the first equation can be used to extract the shear viscosity $\eta$ and relaxation time $\tau_\text{shear}$ and the second relates the bulk viscosity and relaxation time to the imaginary part of the trace components. 
In the limit $d=4$ we recover the well known relation that is for exampled used in refs.~\cite{Czajka:2017,Kovtun_2012} up to a prefactor in the first term of \cref{eq:kubo_formula_zeta_kovtun} when compared with~\cite{Kovtun_2012}.
However, our result coincides with the static susceptibilities \cref{eq:response_order_zero} for $d$ dimensions. 
We furthermore do also find Kubo formulas where the order of limits of the frequency and spatial momentum are exchanged. 
This has been shown previously in ref~\cite{Jeon:2025}.  
More generally, we find for arbitrary spatial momentum in $d$ spacetime dimensions
\begin{equation}
    \begin{split}
       & \frac{\iu }{6 }\lim_{\omega\rightarrow 0 }\lim_{|\vec{k}|\rightarrow 0 }\Delta_{ijlm} \frac{\partial}{\partial \omega} G^{ijlm}(\omega,\vec{k}) = -\frac{\eta(d-2)(d-1)}{12}, \\
       & -\iu \lim_{|\vec{k}|\rightarrow 0}\lim_{\omega\rightarrow 0} \frac{\bar\epsilon + \bar p }{\vec{k}^2 \frac{\partial}{\partial\omega} \hat \Delta_\nu^i\hat \Delta_\beta^i G_\mathrm{R}^{00\alpha\beta}(\omega,\vec{k})} = \frac{\eta }{\bar \epsilon + \bar p }, \\
       & \left( \lim_{|\vec{k}|\rightarrow 0 }\lim_{\omega\rightarrow 0 } \frac{2(\bar\epsilon + \bar p )}{\vec{k}^4 \frac{\partial^2}{\partial\omega^2} \hat\Delta_\nu^i\hat\Delta_\beta^i G_\mathrm{R}^{0 \nu 0 \beta}(\omega,\vec{k})} \right)^{1/2} = \frac{\eta }{\bar \epsilon + \bar p }, \\
       & \lim_{|\vec{k}|\rightarrow 0}\lim_{\omega\rightarrow 0} \frac{\partial}{\partial\omega} G_\mathrm{R}^{0000}(\omega,\vec{k}) = \frac{\iu(2(d-2)\eta + (d-1)\zeta )}{c_\mathrm{S}^4(d-1)}, \\
       & \lim_{|\vec{k}|\rightarrow 0}\lim_{\omega\rightarrow 0} \frac{\partial}{\partial\omega} \delta_{mn}G_\mathrm{R}^{00 mn}(\omega,\vec{k}) = \frac{2\iu(d-2)\eta }{c_\mathrm{S}^2}, \\
       & \lim_{|\vec{k}|\rightarrow 0}\lim_{\omega\rightarrow 0} \frac{\partial}{\partial\omega} \delta_{ij}\delta_{mn}G_\mathrm{R}^{ijmn}(\omega,\vec{k}) = 2\iu(d-2)(d-1)\eta, \\
       & \lim_{|\vec{k}|\rightarrow 0}\lim_{\omega\rightarrow 0} \frac{\partial}{\partial\omega} G_\mathrm{R}^{\Delta P \Delta P}(\omega,\vec{k}) = \iu\zeta,
    \end{split}
\end{equation}
where 
\begin{equation}
\Delta P = (d-1)^{-1}\sum_{j=1}^{d-1} T^j_j - c_\mathrm{S}^{-2} \epsilon,
\end{equation}
is the pressure operator as defined in ref.~\cite{Jeon:2025}.
The Kubo formulae are in accordance with the results in~\cite{Jeon:2025} if one takes $d=4$, up to a global sign which stems from a difference in conventions.

\subsection{Relation to other correlation functions}
While the retarded response function specified through \cref{eq:responsefunction_spatialspatial,eq:responsefunction_timespatial,eq:responsefunction_timetime} can be used to directly access thermodynamic properties or Kubo formulas as discussed above, it is also directly related to further correlation functions of the energy-momentum tensor. 
Among these are two correlation functions at fixed order, and the spectral correlation functions. 
The former two are defined by 
\begin{equation}
    \begin{split}
        &G_{+}^{\mu\nu\alpha\beta}(x-y) = \frac{1}{Z}\text{Tr}\left\{ e^{-\beta \hat H}  \hat T^{\mu\nu}(x) \hat T^{\alpha\beta}(y)\right\} \\
        &= \int \frac{\dd \omega \dd^{d-1} \vec{k} }{(2\pi)^d}\eu^{-\iu\omega(x^0-y^0) + \iu\vec{k}(\vec{x}-\vec{y})}G_{+}^{\mu\nu\alpha\beta}(\omega,\vec{k}),
    \end{split}
    \label{eq:def_posfrequency_correlation}
\end{equation}
and
\begin{equation}
    \begin{split}
        &G_{-}^{\mu\nu\alpha\beta}(x-y) = \frac{1}{Z}\text{Tr}\left\{ e^{-\beta \hat H}  \hat T^{\alpha\beta}(y) \hat T^{\mu\nu}(x) \right\}  \\
        &= \int \frac{\dd \omega \dd^{d-1} \vec{k} }{(2\pi)^d}\eu^{-\iu\omega(x^0-y^0) + \iu\vec{k}(\vec{x}-\vec{y})}G_{-}^{\mu\nu\alpha\beta}(\omega,\vec{k}),
    \end{split}
    \label{eq:def_negfrequency_correlation}
\end{equation}
where $\hat H$ is the Hamiltonian, $Z$ the canonical partition function, and $\hat T^{\mu\nu}(x)$ denote operators in the Heisenberg picture.

Because the position space correlation functions are complex conjugates of each other, $G_{\pm}^{\mu\nu\alpha\beta}(x-y) = G_{\mp}^{\mu\nu\alpha\beta}(x-y)^*$, one has in Fourier space $G_{\pm}^{\mu\nu\alpha\beta}(\omega, \mathbf{k}) = G_{\mp}^{\mu\nu\alpha\beta}(-\omega, -\mathbf{k})^*$. We also note the trivial permutation relations $G_{+}^{\mu\nu\alpha\beta}(x-y) = G_{-}^{\alpha\beta\mu\nu}(y-x)$, or in Fourier space, $G_{+}^{\mu\nu\alpha\beta}(\omega, \mathbf{k}) = G_{-}^{\alpha\beta\mu\nu}(-\omega, - \mathbf{k})$. 
KMS symmetry~\cite{Kubo:1957mj,Martin:1959jp} implies the more interesting relation $G_{+}^{\mu\nu\alpha\beta}(x^0-y^0, \mathbf{x}-\mathbf{y}) = G_{-}^{\mu\nu\alpha\beta}(x^0-y^0-\iu \beta, \mathbf{x}-\mathbf{y})$, or in Fourier space, $G_{+}^{\mu\nu\alpha\beta}(\omega, \mathbf{k}) = \eu^{\beta \omega}G_{-}^{\mu\nu\alpha\beta}(\omega, \mathbf{k})$.

The spectral correlation function is defined as the expectation value of the commutator
\begin{equation}
    \begin{split}
        &\rho^{\mu\nu\alpha\beta}(x-y) = \frac{1}{Z}\text{Tr}\left\{ e^{-\beta \hat H} \left[ \hat T^{\mu\nu}(x), \hat T^{\alpha\beta}(y) \right]   \right\} \\
        &= \int \frac{\dd \omega \dd^{d-1} \vec{k} }{(2\pi)^d}\eu^{-\iu\omega(x^0-y^0) + \iu\vec{k}(\vec{x}-\vec{y})}\rho^{\mu\nu\alpha\beta}(\omega,\vec{k}).
    \end{split}
    \label{eq:def_spectral_correlation}
\end{equation}
We note that 
\begin{equation}
\begin{split}
    \rho^{\mu\nu\alpha\beta}(\omega,\vec{k}) = & G_{+}^{\mu\nu\alpha\beta}(\omega,\vec{k}) - G_{-}^{\mu\nu\alpha\beta}(\omega,\vec{k}) \\
    = & \left(\eu^{\beta \omega } - 1 \right)G_{-}^{\mu\nu\alpha\beta}(\omega,\vec{k}).
\end{split}\label{eq:rhoG-Relation}
\end{equation}
In position space, $\smash{\rho^{\mu\nu\alpha\beta}(x-y)}$ is purely imaginary, which implies in Fourier space $\rho^{\mu\nu\alpha\beta}(\omega,\vec{k}) = -\rho^{\mu\nu\alpha\beta}(-\omega,-\vec{k})^*$. Moreover, in a time-reflection and parity symmetric situation one has also $\rho^{\mu\nu\alpha\beta}(\omega,\vec{k}) =-\rho^{\mu\nu\alpha\beta}(-\omega,-\vec{k})$ and therefore a real spectral function, $\rho^{\mu\nu\alpha\beta}(\omega,\vec{k}) \in \mathbb{R}$. There is also a spectral representation for $G_R^{\mu\nu\alpha\beta}(\omega, \mathbf{k})$ in terms of $\rho^{\mu\nu\alpha\beta}(\omega, \mathbf{k})$ from which it follows that the spectral function is given by imaginary part of the retarded correlation function, $\rho^{\mu\nu\alpha\beta}(\omega,\vec{k})=2\, \mathrm{Im}\,G_\mathrm{R}^{\mu\nu\alpha\beta}(\omega,\vec{k})$. 
Together with \eqref{eq:rhoG-Relation} this implies
\begin{equation}
    G_{-}^{\mu\nu\alpha\beta}(\omega,\vec{k}) =2  \mathrm{n}_\mathrm{B}(\omega) \, \mathrm{Im}\,  G_\mathrm{R}^{\mu\nu\alpha\beta}(\omega,\vec{k}),
    \label{eq:relation_gplus_gr}
\end{equation}
with the Bose function $\smash{\mathrm{n}_\mathrm{B}(z) = 1/(\eu^{\beta z } - 1 )}$.
A more detailed overview of the different two-point correlation functions can be found in refs.~\cite{Calzetta:2008iqa, Bellac:2011kqa, Kapusta_Gale_2006}.

%% file: inputs/prod_and_decay.tex
Based on our calculation of the response function, we will now consider two applications to fluids interacting with Einstein gravity.
First, the gravitational wave spectrum produced by thermal fluctuations, and second, an application of Mueller-Israel-Stewart theory to gravitational wave damping in an expanding universe.

\subsection{Gravitational wave production}
The linear response \cref{eq:linear_response_definition} suggests that a fluid with fluctuations around thermal equilibrium will  generate gravitational waves~\cite{Weinberg:2008zzc}. 
This assumes that the latter are weakly interacting with the fluid and can escape, instead of being in detailed balance with the fluid themselves.
This has already been considered in ref.~\cite{Ghiglieri:2020mhm}, while ref.~\cite{Drewes:2023oxg} has considered this problem on an expanding Friedmann-Lemaître-Robertson-Walker (FLRW) background. 
The production rate of gravitational waves is a key ingredient for the calculation of the spectrum and energy density of gravitational radiation.

Assuming fluctuations are small, we can utilize the linear response relation and determine the production rate at small wave numbers completely from the fluid dynamic equations. 
In both former mentioned works, the rate $\Pi_\mathrm{GW}(\omega,\vec{k})$ per unit volume was found to be given by 
\begin{equation}
    \begin{split}
      &\Pi_\mathrm{GW}(\omega,\vec{k}) =\\
      &\qquad\,\frac{1}{2}\mathcal{P}_{\mathrm{(TT)}}^{ijmn}\int_{t,\vec{x}} e^{\iu(\omega t - \vec{k}\vec{x})} \langle T_{ij}(0,\vec{0}) T_{mn}(t,\vec{x})\rangle,
    \end{split}
    \label{eq:gw_spectrum_definition}
\end{equation}
where $\mathcal{P}_{\mathrm{TT}}$ is the projector onto the transverse traceless modes, defined by \cref{eq:projectors_orthogonal_subspaces}.
For a textbook calculation we refer the reader to refs~\cite{Bellac:2011kqa,Kapusta_Gale_2006}.

\begin{figure}[htbp] 
    \centering
    \includegraphics[width=1.0\columnwidth]{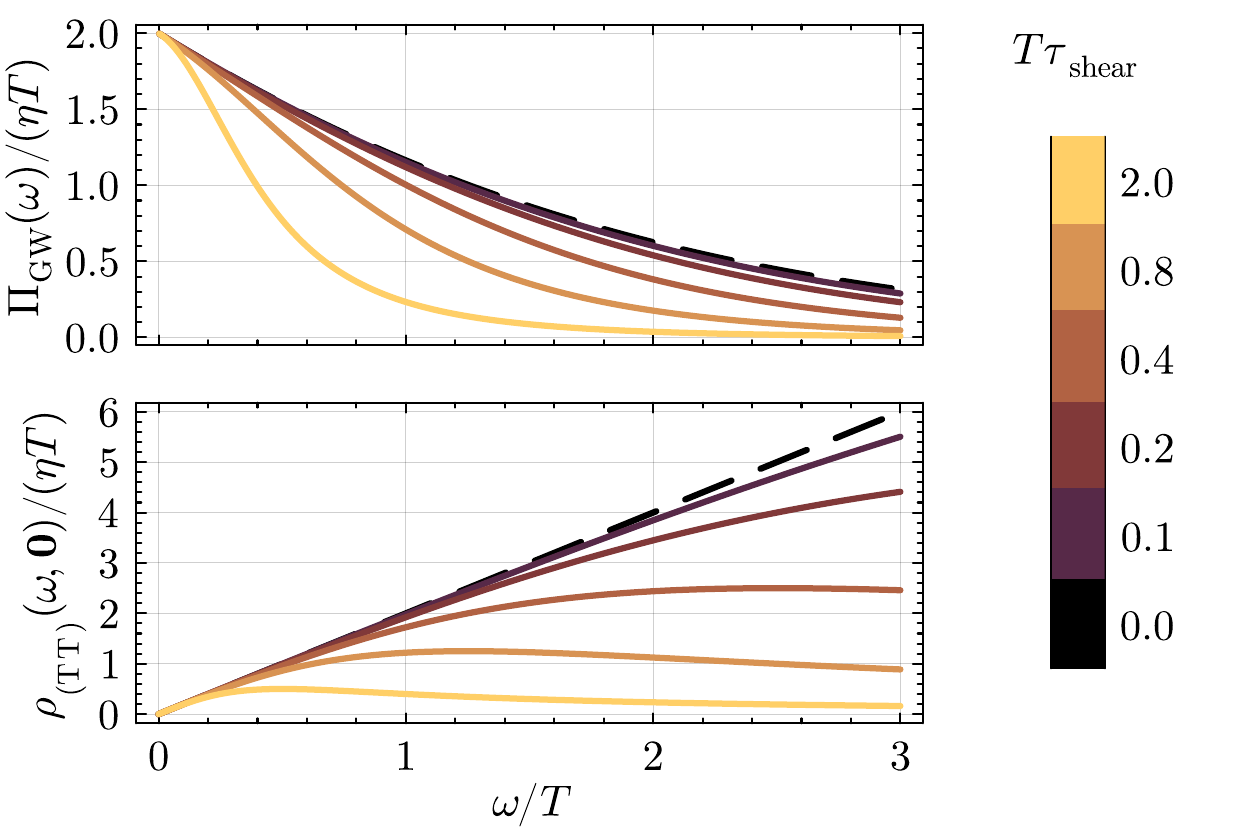}
    \caption{
    Gravitational wave production rate $\Pi_\mathrm{GW}(\omega)$ (upper panel) and transverse part of the spectral function with $\smash{\rho_\mathrm{(TT)}(\omega,\vec{k})=\mathcal{P}_\mathrm{(TT)\,ijmn} \rho^{ijmn}(\omega,\vec{k})}$ (lower panel) in dimensionless units over $\omega/T\in[0,5]$ at different values of $T\tau_\mathrm{shear}$.
    The first-order limit is shown as the black dashed line where $\tau_\mathrm{shear}=0$.}
    \label{fig:production_rate_over_t4}
\end{figure}

The production rate is the Fourier transform of the transverse traceless part of the correlation function defined in \cref{eq:def_negfrequency_correlation}.
Taking \cref{eq:responsefunction_spatialspatial} and making use of \cref{eq:relation_gplus_gr} we can calculate the gravitational wave production rate $\Pi_\mathrm{GW}(\omega,\vec{k})$ within the second order Mueller-Israel-Stewart theory. 
This leads to 
\begin{equation}
\begin{split}
      \Pi(\omega,\vec{k}) =& 2\mathrm{n}_\mathrm{B}(\omega)\mathrm{Im}\frac{b_1 }{d_1}=n_\mathrm{B}(\omega )\frac{2\eta \omega }{1 + \omega^2 \tau_\text{shear}^2}.
\end{split}
    \label{eq:gw_spectrum_hydro}
\end{equation}
The result is also shown in \cref{fig:production_rate_over_t4} in dimensionless form.

Compared to the first order results given by the limit $\tau_\text{shear}\rightarrow 0$, the rate is stronger suppressed the larger the frequency with respect to the temperature gets. 
Furthermore, while the rate still scales linearly in the shear viscosity coefficient, a greater relaxation time does significantly suppress the emitted spectrum even at low frequencies in contrast to the first order limit.

\subsection{Damping of gravitational waves}
Another interesting effect which appears in the context of gravitational fields coupled to a fluid is the dissipation of a gravitational wave's energy when passing through a viscous medium. 
Tensor perturbations in the fluid mix with those of the metric.
In a cosmological setting, the effect depends on the expansion of the universe as well as on the shear viscosity and corresponding relaxation time.
In astrophysical situations, gravitational waves can scatter at dense, viscous stars~\cite{Boyanov:2024jge}.

For the following consideration we will assume a spatially flat FLRW background with metric given by 
\begin{equation} 
  \dd s^2 = \bar g_{\mu\nu}\dd x^\mu \dd x^\nu = a^2(\tau)\left[-\dd \tau^2 +\delta_{ij}\dd x^i \dd x^j\right],
  \label{eq:flrw_metric}
\end{equation}
with expansion scalar $a(\tau)$, conformal time $\tau$ defined so that $\dd \tau = \dd t / a(\tau)$, and comoving coordinates $\mathbf{x}$ for $d=4$ spacetime dimensions. 

Metric perturbations are in general decomposed into
\begin{equation}
  \begin{split}
    \delta g_{00}(x) &= -2a^2(\tau)\psi(x),\\ 
    \delta g_{0i}(x) &= a^2(\tau)w_j(x), \\
    \delta g_{ij}(x) &= 2a^2(\tau)\left[ h_{ij}(x) - \delta_{ij}\phi(x) \right],
  \end{split}
  \label{eq:flrw_fluctuations}
\end{equation}
with the scalars $\psi$ and $\phi$, vector $w_i$ and symmetric transverse traceless tensor part $h_{ij}$, following the notation of~\cite{Ma:1995ey}.
Employing the Poisson gauge, 
\begin{equation}
  \delta^{ij}\nabla_i w_j(x) = 0, \quad \delta^{ik}\nabla_k h_{ij}(x) = 0,
  \label{eq:poisson_gauge}
\end{equation}
leaves us with six remaining degrees of freedom.
For the remainder of this section we shall only consider the dynamics of the two tensor perturbations described by $h_{ij}$.%, which are gauge invariant~\cite{Mukhanov:1990me}. 

A related decomposition into scalar, vector and tensor perturbations can also be done for the fluid fields. We concentrate again on the tensor perturbations, and write
\begin{equation}
  \delta \pi_{ij}(x) = a(\tau)^2 \hat \pi_{ij}(x),
\end{equation}
where $\hat \pi_{ij}$ is traceless, $\delta^{ij}\hat \pi_{ij}=0$, and satisfies $\delta^{ij}\partial_i \hat\pi_{jk}=0$.

Spatial homogeneity of the background allows us to consider the Fourier representation of these modes where each amplitude is proportional to $\eu^{\iu\vec{q}\vec{x}}$.
The Einstein field equations,
\begin{equation}
  R^{\mu\nu} - \frac{R}{2}g^{\mu\nu} = 8\pi G_\mathrm{N} T^{\mu\nu},
  \label{eq:einstein_field_equation}
\end{equation}
couple small perturbations of the metric to the ones of the fluid. 
The linearized Einstein tensor and the Christoffel symbols up to first order can be found in~\cite{Weinberg:2008zzc}.
Using these together with the fluids equations of motion \cref{eq:energy_linearized_general_coords,eq:velocity_linearized_general_coords,eq:bulk_pressure_linearized_general_coords,eq:shear_linearized_general_coords} yields the evolution equation of the tensor modes which is given by
\begin{equation}
    -\frac{\partial^2}{\partial\tau^2} h_{ij} - 2 a H \frac{\partial}{\partial\tau} h_{ij} - \vec{q}^2 h_{ij}  =  - 8\pi G_\text{N} a^2 \hat \pi_{ij},
  \label{eq:eom_tensor_perturbations_flrw}
  \end{equation}
with the Hubble rate $H= (1/a^2) (\dd a / \dd \tau) = (1/a) (\dd a/ \dd t)$ and gravitational coupling constant $G_\text{N}$. 

The shear stress perturbation evolves according to 
\begin{equation}
  \frac{\tau_\mathrm{shear} }{a} \frac{\partial}{\partial \tau} \hat \pi_{ij} +  \left( 1+4 H \tau_\mathrm{shear} \right)\hat \pi_{ij} = -\frac{2\eta}{a } \frac{\partial}{\partial\tau} h_{ij}.
  \label{eq:eom_shear_perturbations_flrw}
\end{equation}
We obtain thus two coupled differential equations, one second order and one first order, for the metric perturbations $h_{ij}$ and fluid tensor perturbations $\hat \pi_{ij}$. 

In the limit $\tau_\mathrm{shear} \to 0$, eq.\ \eqref{eq:eom_shear_perturbations_flrw} becomes basically a constraint and one obtains the equation of motion (cf.\ ref.~\cite{Weinberg:1972kfs} )
\begin{equation}
  \begin{split}
    -\frac{\partial^2}{\partial\tau^2} h_{ij} - \left[ 2 a H + 16\pi G_\text{N} a \eta \right]\frac{\partial}{\partial\tau} h_{ij} - \vec{q}^2 h_{ij} = 0. 
  \end{split}
\end{equation}
Gravitational waves are diluted by the cosmological expansion and additionally damped by shear viscous dissipation.

Another interesting limit is the one of a Minkowski space background, $a=1$. We assume that the viscosity $\eta$ and relaxation time $\tau_\mathrm{shear}$ are independent of time, and find by combing eqs. \eqref{eq:eom_tensor_perturbations_flrw} and \eqref{eq:eom_shear_perturbations_flrw}
\begin{equation}
\begin{split}
  & \tau_\text{{shear}}\frac{\partial^3}{\partial t^3} h_{ij} + \frac{\partial^2}{\partial t^2} h_{ij} \\
  & + \left( 16\pi G_\text{N} \eta + \tau_\text{shear} \mathbf{q}^2 \right) \frac{\partial}{\partial t} h_{ij} + \mathbf{q}^2 h_{ik} = 0.
\end{split}
\end{equation}
Writing the perturbations $h_{ij}$ as Fourier modes $\sim \eu^{-i\omega t}$ leads to a dispersion relation 
\begin{figure*}[htbp]
  \centering
  \includegraphics[height=0.45\textheight]{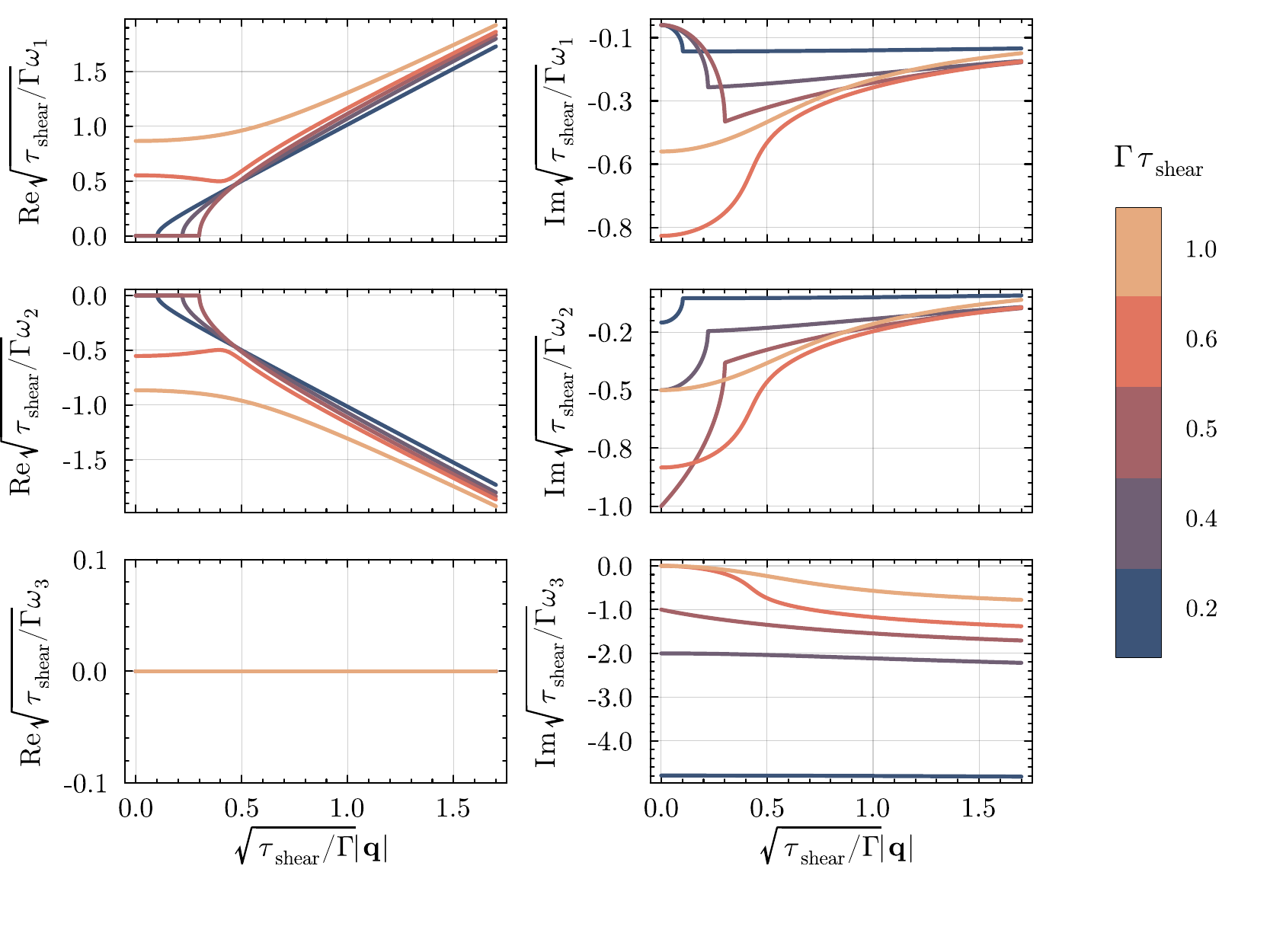}
  \caption{The three dispersion relations obtained from \cref{eq:gw_dispersion} in dimensionless form for different values of $\tau_\text{shear}\Gamma$. 
  The two solutions $\omega_{1,2}$ correspond to the standard gravitational wave solutions.
  The lower row shows the solution which is only present at non-vanishing relaxation time $\tau_\mathrm{shear}$.}
  \label{fig:dispersion_damping_waves}
\end{figure*}

\begin{equation} 
  \iu\tau_\text{shear} \omega^3 - \omega^2 -\iu\omega \left[ \Gamma  +  \tau_\text{shear}\vec{q}^2 \right] + \vec{q}^2 = 0.
  \label{eq:gw_dispersion}
\end{equation}  
with the damping constant $\Gamma=16\pi G_\text{N} \eta$. For $\vec{q}^2 \gg \Gamma/\tau_\text{shear}$ one has the standard gravitational waves with $\omega^2 = \vec{q}^2$, as well as a relaxing shear stress mode with $\omega = -\iu/\tau_\text{shear}$. Another interesting limit is $\vec q^2=0$ where one solution is $\omega=0$ while the other two follow from
\begin{equation}
  \omega^2 + \frac{\iu}{\tau_\text{shear}} \omega - \frac{\Gamma}{\tau_\text{shear}} = 0,
\end{equation}
which is the dispersion relation for a damped harmonic oscillator. Also because $\Gamma\sim G_\text{N}$ is small, one would expect this oscillator to be in the overdamped regime. More generally, for finite $\vec q^2$, the algebraic equation \eqref{eq:gw_dispersion} has three complex solutions $\omega_1$, $\omega_2$ and $\omega_3$ which describe a mixture of gravitational waves and the relaxing shear stress mode.

The real and imaginary parts of these complex solutions are shown in \cref{fig:dispersion_damping_waves}.
While the first two solutions describe ordinary damping of gravitational waves by viscous effects the third solution has a vanishing real part and only contributes a decaying mode in the linear regime. 
In the large wave number limit the solution becomes constant.

%% file: inputs/conclusion.tex
We have discussed here the linear response of relativistic fluids to metric perturbations.

Starting with the energy-momentum tensor and the general setup of response theory we have first re-derived the Ward identity following from diffeomorphism symmetry, recovering results found in~\cite{Czajka:2017,Moore:2010bu,Jeon:2025}.

After introducing the Mueller-Israel-Stewart description of relativistic fluids, we explicitly calculated the static susceptibilities in general coordinates as well as the linearized equations of motion, containing both bulk viscous pressure and shear stress tensor contributions. 
Specifically for a homogeneous fluid on a Minkowski background we furthermore calculated the full response function by inverting the equations of motion. This explicit expression, complied in eqs.\ \eqref{eq:responsefunction_spatialspatial}, \eqref{eq:responsefunction_timespatial} and \eqref{eq:responsefunction_timetime}, can be used as a starting point for further investigations. First, it allows to check the known Kubo formulas and static susceptibilities in $d=4$ spacetime dimensions, but also to extend them to general number of $d$ spacetime dimensions, recovering the results from refs.~\cite{Czajka:2017,Jeon:2025,Moore:2010bu,Kovtun_2012}.

In addition, this result can be used to find the production rate for gravitational waves from thermal fluctuations in a relativistic fluid, in the soft regime, see \cref{eq:gw_spectrum_hydro}. In the limit where the shear stress relaxation time $\tau_\text{shear}$ vanishes, we recover the known result that is linear in the shear viscosity. 
At finite shear stress relaxation time, however, the production rate additionally suppressed for larger frequencies. 

Finally, as a last application we turned to the dissipation of gravitational waves, on a FLRW background, when passing a viscous, relativistic fluid. For vanishing shear stress relaxation time we reproduced the known damping induced by the shear viscosity coefficient as already discussed in ref.~\cite{Weinberg:1972kfs}. Modifications arise when the relaxation time is finite, such as the appearance of an additional excitation mode corresponding to the relaxing shear stress. For intermediate wave numbers these excitations mix with gravitational waves.

The work presented here can be extended in several directions. One possibility is to take additional conserved currents into account, like the one for baryon number following from a global symmetry, or the one for electric charge following from an additional gauge symmetry. This leads to a larger set of equations of motion, but the principle method to invert the equations of motion is applicable there, as well. Another direction can be to study different formulations of relativistic fluid dynamics. It is conceivable that further studies of the interplay between relativistic fluids and spacetime geometry lead to further insights into the dynamics of relativistic fluids.

%% file: inputs/appendix.tex
\section{Tensor Decomposition}
\label{sec:appendix_projectors}
Starting with the shear modes, \cref{eq:shearstress_linearized_tmn}, one can express the different variations of the transverse-traceless projector $\Delta^{ijmn}$, orthogonal projector $\Delta^{ij}$ and momentum $k_n$ in terms of the projectors \cref{eq:projectors_orthogonal_subspaces} such that 
\begin{equation}
    \begin{split}
        \Delta^{ij\alpha\beta} =& \frac{1}{d-1} \mathcal{P}_{\mathrm{(S1)}}^{ij\alpha\beta} + \mathcal{P}_{\mathrm{(TT)}}^{ij\alpha\beta} + \mathcal{P}_{\mathrm{(V)}}^{ij\alpha\beta} \\
        &+ \left( 1-\frac{1}{d - 1} \right)\mathcal{P}_{\mathrm{(S2)}}^{ij\alpha\beta} \\
        &- \frac{1}{d - 1}\left( \hat\Delta^{ij}\hat k^\alpha \hat k^\beta + \hat k^i \hat k^j \hat\Delta^{\alpha\beta} \right)\\
        \Delta^{ijm\alpha} \hat k_m =& \frac{1}{2}\left( \hat k^i \hat\Delta^{j\alpha} + \hat k^j\hat\Delta^{i\alpha} \right) \\
        &+ \left( 1-\frac{1}{d-1}  \right)\hat k^i \hat k^j \hat k^\alpha - \frac{1}{d-1}\hat\Delta^{ij}\hat k^\alpha \\
        \Delta^{ijmn}\hat k_m \hat k_n =& \left( 1-\frac{1}{d-1} \right)\hat k^i \hat k^j -\frac{1 }{d-1}\hat\Delta^{ij} .
    \end{split}
\label{eq:shear_building_blocks_projected}
\end{equation}
Inserting \cref{eq:shear_building_blocks_projected} into the definition of $D^{ijmn}$ this leads to 
\begin{equation}
    \begin{split}
        D^{ijmn} =& \frac{d_1 }{d-1 }\mathcal{P}_\mathrm{(S1)}^{ijmn} + \left( 1-\frac{1 }{d-1} \right)(d_1 + 2 d_2 )\mathcal{P}_{\mathrm{(S2)}}^{ijmn} \\
        +& d_1 \mathcal{P}_{\mathrm{(TT)}}^{ijmn} + (d_1 + d_2) \mathcal{P}_{\mathrm{(V)}}^{ijmn} \\
        &- \frac{d_1 }{d-1}\hat k^i \hat k^j \hat \Delta^{mn} - \frac{d_1 + 2d_2 }{d-1}\hat \Delta^{ij}\hat k^m \hat k^n.
    \end{split}
\end{equation}
The previous equation shows that the right-hand side of the shear stress equation of motion is split into the three separate parts 
\begin{equation}
    \begin{split}
        D_\mathrm{(TT)}^{ijmn} =& d_1 \mathcal{P}_{\mathrm{(TT)}}^{ijmn} \\
        D_\mathrm{(V)}^{ijmn} =& (d_1 + d_2) \mathcal{P}_{\mathrm{(V)}}^{ijmn} \\
        D_\mathrm{(S)}^{ijmn} =& \frac{d_1 }{d-1} \mathcal{P}_{\mathrm{(S1)}}^{ijmn} + \left( 1-\frac{1}{d-1} \right)(d_1 + 2 d_2)\mathcal{P}_{\mathrm{(S2)}}^{ijmn} \\
        &- \frac{d_1 }{d-1}\hat k^i \hat k^j \hat\Delta^{mn} - \frac{d_1 + 2 d_2 }{d-1}\hat \Delta^{ij}\hat k^m \hat k^n .
    \end{split}
    \label{eq:lhs_shear_equation_tensor_decomposition}
\end{equation}
The first part maps the transverse traceless parts onto themselves and the second part maps vectors onto vectors. 
The remainder term $\smash{D_\mathrm{(S)}^{ijmn}}$ mixes the two scalars with each other. 

The right-hand side of equation \cref{eq:shearstress_linearized_tmn} is a bit more involved but can be written down as 
\begin{equation}
    \begin{split}
        B^{ij\alpha\beta} =& \left[ \frac{b_1 }{d-1 } + b_4 \left( 1 - \frac{1}{d-1 } \right) \right]\mathcal{P}_\mathrm{(S1)}^{ij\alpha\beta} \\
        +& \left( 1- \frac{1}{d-1 } \right)(b_1 + b_4 ) \mathcal{P}^{ij\alpha\beta}_\mathrm{S2} \\
        +& (b_1 + b_4)\mathcal{P}_\mathrm{(V)}^{ij\alpha\beta} + b_1 \mathcal{P}_\mathrm{(TT)}^{ij\alpha\beta} \\
        +& \frac{b_2 }{2}\left( \hat k^i \hat \Delta^{i\alpha} u^\beta + \hat k^j \hat \Delta^{i\alpha} u^\beta + \hat k^i \hat \Delta^{j\beta} u^\alpha \right. \\
        +& \left.\hat k^j\hat \Delta^{i\beta} u^\alpha\right) + \hat k^i \hat k^j \left( 1-\frac{1}{d-1 } \right)\times\\
         &\bigg[b_2\left( \hat k^\alpha u^\beta + u^\alpha \hat k^\beta  \right) + b_3 u^\alpha u^\beta \\
        -& \left( \frac{b_1}{d-2} + b_4 \right)\hat\Delta^{\alpha\beta}\bigg] \\
        -&\frac{\hat\Delta^{ij}}{d-1}\times\bigg[b_2\left( \hat k^\alpha u^\beta + u^\alpha \hat k^\beta  \right)\\
        +&b_3 u^\alpha u^\beta + \left( b_1 + b_4 \right)\hat k^\alpha\hat k^\beta \bigg].
    \end{split} 
\end{equation} 
The $B$-tensor can also be split into the three distinct subspaces (TT), (V) and (S) with $\smash{B^{ij\alpha\beta} = B_{\mathrm{(TT)}}^{ij\alpha\beta} + B_{\mathrm{(V)}}^{ij\alpha\beta} + B_\mathrm{(S)}^{ij\alpha\beta}}$ where the different parts are given by 
\begin{equation}
    \begin{split}
        B_{\mathrm{(TT)}}^{ij\alpha\beta} =& \mathcal{P}^{ij}_\mathrm{(TT)mn}B^{mn\alpha\beta} = b_1 \mathcal{P}_{\mathrm{(TT)}}^{ij\alpha\beta}, \\
        B_{\mathrm{(V)}}^{ij\alpha\beta} =& \mathcal{P}^{ij}_\mathrm{(V)mn}B^{mn\alpha\beta},\\
        =&\left( b_1 + b_4  \right)\mathcal{P}_{\mathrm{(V)}}^{ij\alpha\beta} + \frac{b_2}{2}\left( \hat k^i \hat \Delta^{j\alpha}u^\beta \right.\\
        +&\left.\hat k^j\hat\Delta^{i\alpha} u^\beta + \hat k^i \hat \Delta^{j\beta}u^\alpha + \hat k^j \hat \Delta^{i\beta}u^\alpha \right),\\
        B_{\mathrm{(S)}}^{ij\alpha\beta} =& \left[ \frac{b_1 }{d-1 } + b_4 \left( 1 - \frac{1}{d-1 } \right) \right]\mathcal{P}_\mathrm{(S1)}^{ij\alpha\beta} \\
        +& \left( 1- \frac{1}{d-1 } \right)(b_1 + b_4 ) \mathcal{P}^{ij\alpha\beta}_\mathrm{S2} \\
        +& \hat k^i \hat k^j \left( 1-\frac{1}{d-1 } \right)\times \bigg[b_2\left( \hat k^\alpha u^\beta + u^\alpha \hat k^\beta  \right) \\
        +& b_3 u^\alpha u^\beta - \left( \frac{b_1}{d-2} + b_4 \right)\hat\Delta^{\alpha\beta}\bigg] \\
        -&\frac{\hat\Delta^{ij}}{d-1}\times\bigg[b_2\left( \hat k^\alpha u^\beta + u^\alpha \hat k^\beta  \right)\\
        +&b_3 u^\alpha u^\beta + \left( b_1 + b_4 \right)\hat k^\alpha\hat k^\beta \bigg].
    \end{split}
    \label{eq:rhs_shear_equation_tensor_decomposition}
\end{equation}